# Constraints on Interpretations of Quantum Mechanics

Casey Blood
Professor Emeritus of Physics, Rutgers University
Sarasota, FL
Email: CaseyBlood@gmail.com

## Abstract

A succinct statement and justification of all the principles necessary to understand and evaluate interpretations of quantum mechanics is given. These principles provide strong constraints on interpretations. They imply the particle-like properties of mass, energy, momentum, spin, charge, and locality are actually properties of the wave function; and this in turn implies there is no evidence for the existence of particles. In addition, there is currently no experimental evidence for collapse, and a theory of collapse encounters significant hurdles. Further, the probability law is found to rule out the many-worlds interpretation, so all three major interpretations encounter serious to fatal problems. An interpretation which conforms to all the principles is given.

PACS numbers: 03.65-w, 03.65.Ta

## Introduction.

Quantum mechanics is a highly successful theory which accurately describes physical reality in a wide range of situations. Its use of the non-intuitive wave function, however, means that the true nature of the physical reality described is far from clear. Because of this, many different interpretations—conceptual pictures of the nature of the underlying reality—have been proposed. There are three major interpretations and perhaps a dozen others, so one sometimes gets the impression that it is not possible to gain a deeper understanding of what underlies the theory.

I think, however, that the situation is not as chaotic as it appears. The reason there are so many interpretations is that their originators were not always aware of all the relevant properties of the theory itself. So what we do here is list and justify all the properties of quantum mechanics relevant to its interpretation. This approach brings order to the interpretive problem and narrows down the set of possible interpretations.

**History.** To understand how a successful theory could require an interpretation, we need to briefly review the history of physics. Classical mechanics, discovered almost entirely by Newton around 1700, was a very successful mathematical scheme for predicting the paths of the atoms that matter was presumably made of. Since it involved the flight of easily visualized particles through our familiar three-dimensional space, classical physics gave a picture of reality which was compatible with our conventional, everyday view of the world. Quantum mechanics, however, is a different story. It was discovered in the search for a mathematical scheme which would explain classically unsolvable problems such as why the hydrogen atom radiates certain colors of light. Rather than being the brainchild of a single person, however, there were at least half a dozen major physicists who, over a period of 25 years, each discovered an essential piece of the puzzle. And with each piece, the mathematics moved farther away from our everyday view of the world.

The end result of these separate insights—the equation discovered by Schrödinger in 1926—was a scheme which is highly successful. In addition to the hydrogen atom spectrum, quantum mechanics describes a great many phenomena—all the properties of atoms and nuclei, semiconductors, lasers, the systematics of elementary particles—and it has never given a result in conflict with experiment. But its mathematics involves the mysterious wave function rather than the easily-visualized particles, so it no longer has a clear correspondence with our commonplace view of reality. Because of this, one needs an *interpretation* of the mathematics to understand how it relates to our perceived physical world.

**Peculiarities of Quantum Mechanics.** Quantum mechanics is difficult to understand because it has two peculiar properties. The first, illustrated by the Schrödinger's cat experiment in the next section, is that the wave function of quantum mechanics contains several *simultaneously existing* versions of reality—the cat is both alive and dead at the same time. We perceive only one of those possible versions of reality. But quantum mechanics treats all versions equally, so it does not tell us why we perceive a *particular* version.

The fact that there are several versions of reality leads to the second peculiar property, probability. Through an empirically verified law which appears to be separate from the rest of the mathematics, quantum mechanics tells us the *probability* of seeing a live cat or a dead cat. There is currently no clear understanding of this 'rolling-the-dice' aspect of the theory. (There is one other peculiar quantum mechanical property, non-locality—see appendix A8. But, although it is of interest in its own right, non-locality is not a major factor in determining the correct interpretation.)

**Major interpretations.** There are three major interpretations which attempt to give an understanding of these properties. The first is to suppose that, in addition to the many quantum versions of reality, there is an actual, objective reality made of particles, and it is the particles, rather than the wave functions, that we perceive. A second potential way out of the multiple-versions-of-reality problem is to suppose the wave function 'collapses' down to just one version; the dead cat wave function might collapse to zero, for example, leaving just a live cat (wave function) to be perceived. And a third possible solution, the Everett many-worlds interpretation [1], is to suppose all versions always exist, so there are many version of each of us!? We will explain these in more detail and show that each runs into severe difficulties.

**Conclusion.** The implication of these results is that the theory of the physical universe, as it currently stands, is *incomplete*. For if the probability law is to hold, there must be some 'mechanism' that singles out just one version as the one corresponding to our perceptions. But there is no evidence for, and substantial evidence against, the conventionally proposed mechanisms—particles, hidden variables and collapse. Thus the most likely candidate for the singling out mechanism at this time is a perceiving "Mind" not subject to the laws of quantum mechanics (in contrast to the physical brain, which is subject to those laws).

**Primary strategy.** In an attempt to eliminate any pre-judgments on the true nature of physical reality, we use the strategy of examining the connections between the mathematics of quantum mechanics and our *perceptions*.

**Level of the paper.** As much as we could, we took care to make the main text understandable to the general reader, and even to make parts of it a primer on the interpretation of quantum mechanics. The more complicated mathematical derivations of the principles, as well as brief descriptions and analyses of relevant experiments, are relegated to the appendices.

## Contents

# 1. Versions of Reality.

**A. Visualizing the wave function.** The Schrödinger equation of quantum mechanics governs the motion of the *wave functions*, so it is (on the least abstract level) the wave functions, rather than particles, which are the 'physical objects' in the mathematics of quantum mechanics. A useful visual picture of the wave function is that it is matter spread out in a mist or cloud of varying density. The Schrödinger equation determines the shape of the cloud, how it moves through space, and how it responds to other clouds corresponding to other 'particles.' The wave function of a macroscopic object like a cat or a human being, composed of billions of individual wave functions, is extremely complicated, but that does not prevent us from deducing certain relevant general characteristics.

**B. Notation and terminology.** We often have to refer to the 'state' of a particle or detector or observer—the location of a silver atom, a detector reading yes or no, an observer perceiving a live or dead cat. We will sometimes use square brackets, [*label*], and sometimes the 'ket' notation $|label\rangle$, where the *label* describes the relevant property—the position of a silver atom and so on—of the wave function.

> [Technically the ket refers to the abstract 'state vector.' But there is no easy visualization of the state vector, so, since the wave function contains all the necessary information, we will not use the term.]

**C. Many versions of reality.** The world around us certainly appears to be unique, *the* physical world, upon which we all agree. But in quantum mechanics, the highly successful mathematical description of nature, there is no unique physical world; instead there are many simultaneously existing *versions* of physical reality. We will give two examples, nuclear decay and Schrödinger's cat, here. Two other examples of simultaneously existing versions of reality, involving polarization of light and the spin of particles, are given in appendices A1 and A2. These two phenomena are important because they are often used in theoretical arguments and in the experiments employed to test the peculiarities of quantum mechanics.

**D. Nuclear decay.** We first consider the wave function of a single radioactive nucleus. In the mathematics of quantum mechanics, there is both a part of the wave function corresponding to an undecayed nucleus *and* a part corresponding to a decayed (spontaneously split into several parts) nucleus. That is, the state of the system at a given time is

[the nucleus radioactively decays]
**and, simultaneously**
[the same nucleus does not decay]

Both options, both *branches* of the wave function, exist simultaneously! One cannot get around this; the successes of QM depend on it.

Note that we are not saying at this point whether 'the nucleus itself' is both decayed and undecayed. We are simply saying here that *in the mathematics*, 'the nucleus' is both decayed and undecayed. (But in section 7 we will claim there is no such thing as 'the nucleus.')

**E. Schrödinger's cat.** The Schrödinger's cat experiment [2] is a clever and dramatic way of boosting this strange multi-reality situation from the atomic (nuclear) to the macroscopic level. A cat is put in a box along with a vial of cyanide. Outside the box are a radioactive source and a detector of the radiation. The detector is turned on for five seconds and then turned off. If it records one or more counts of radiation, an electrical signal is sent to the box, the vial of cyanide is broken, and the cat dies. If it records no counts, nothing happens and the cat lives.

Classically, there is no problem here (unless you are a cat lover). Either a nucleus radioactively decays, the cat dies and you perceive a dead cat when you open the box; or no nucleus decays, the cat lives and you perceive a live cat. Schematically, in the classical case,

**either**
[nucleus decays] [cat dies] [you perceive a dead cat]
**or**
[no nucleus decays] [cat lives] [you perceive a live cat]

But this is not what happens in the quantum case. There, the wave function of the nucleus, the cat, and you (as the observer) is

[nucleus decays]
[cat dies]
[version 1 of you perceives a dead cat]

**and, simultaneously**
[nucleus does not decay]
[cat lives]
[version 2 of you perceives a live cat]

There are now two full-blown, simultaneously existing versions of physical reality. In one, there is a dead version of the cat, in the other there is a live version. In one, version 1 of you perceives a dead cat, in the other, version 2 of you perceives a live cat. So there are *two simultaneously existing, equally valid versions of you*, each perceiving something different!

In reality, your perceptions will correspond to one version or the other (but that does not necessarily imply that only one version 'exists').

# 2. Quantum Principles of Perception.

We will show here that quantum mechanics does not conflict with our perception of a seemingly classical world, where by "quantum mechanics," we mean the QMA scheme of section 4—no particles, no collapse, no sentient beings, just the linear Schrödinger equation for the wave function.

**A. Agreement of quantum mechanics with our perceptions.**
Schrödinger's cat gives an illustration of a set of general properties which follow from the agreement in all known cases between observation and the quantum mechanically predicted characteristics of the versions of reality:

**2-1.** Quantum mechanics gives many *potential* versions of reality.
**2-2.** In all instances where the calculations and observations can be done, there is always one and only one version whose characteristics corresponds exactly—qualitatively and quantitatively—to our physical perceptions.
**2-3.** Quantum mechanics does not single out any version for perception.
**2-4.** All observers agree on the perceived version.

Property **2-2** is, in my opinion, the pivotal observation in deducing the 'correct' interpretation of quantum mechanics. Although a good deal of work needs to be done to rigorously show this, it would appear to imply that physical reality is composed of wave functions alone, and the wave functions are what 'we' perceive.

**B. Versions are in isolated universes.** The most important mathematical property of quantum mechanics is that it is a linear theory. Because of this, when a wave function divides into a sum of different versions, each version evolves in time *entirely independently* of the other versions present. It is as if (see appendix A3 for a proof)

**2-5.** Each version of reality corresponds to a different, isolated universe. There can be no communication of any kind—via light, sound, touch, email, and so on—between the different versions.

**C. Classical consistency of observation.** (See appendix A4 for details.) The tendency is to suppose that, because quantum mechanics contains several simultaneously existing versions of reality, it must be that in the un-amended theory (no collapse) we should *perceive* several simultaneous versions of reality, akin to a double exposure. But that is not correct. In quantum mechanics proper (QMA), *only the versions of the observer perceive*. Each version, however, is in a separate, isolated universe and can therefore perceive only what is in that universe; a version in one universe can never perceive what occurs in a separate universe. Thus

**2-6.** Perception of more than one outcome of an experiment by an observer can never occur within quantum mechanics.

Next, if we look in the box twice after doing the Schrödinger's cat experiment, we certainly expect, from experience, to see the same thing both times. One can show that this holds in quantum mechanics also.

**2-7.** If two observations are made successively, quantum mechanics implies the same consistency of results as one obtains in a classical universe.

**2-8.** A technical version of **2-7** is that if one *measures* the same property twice in a row, quantum mechanics implies one will get the same result.

Finally, in addition to an observer perceiving only one version of reality, we also know experientially that two observers never disagree on what they perceive. But this also follows from quantum mechanics.

**2-9.** Quantum mechanics implies that two observers can never disagree on what they perceive.

Principles **2-5** through **2-9** constitute much of the content of what is called *measurement theory* [3-5]. (There is also a principle about perceiving only the eigenvalues of relevant operators, but that is not relevant here.) They show that in all circumstances, quantum mechanics itself implies a 'classical' consistency of perceptual results even though no classical (that is, single-version-of-reality) reality was assumed.

**Superselection rule.** Principles **2-5** through **2-9** also imply it is appropriate to apply a 'superselection rule' to the various versions of reality. Linear combinations of, say, the various states of the observer are not prohibited, but (as in the case where one has one universe with charge 2*e* and another with charge 3*e*) when there is no interaction between different solutions of the Schrödinger equation, nothing—no new physical insight—is gained by considering linear combinations of the states. Thus the 'preferred basis problem,' in which linear combinations of versions of the observer are considered, seems to be a red herring.

## 3. Randomness and Probability.

**A. Randomness in a Modified Schrödinger's cat experiment.** We wish to run this experiment many times, so, to avoid killing so many cats, we will modify it. Instead of the detector being hooked up to the box with the cat inside, it is connected to a memory device. Each time the experiment is run (the detector is turned on for 5 seconds and then turned off), the device records a 1 if a decay is detected and a 0 if no decay is detected. On a given run, quantum mechanics does not tell us whether a 0 or a 1 will appear.

**B. Probability.** Suppose we run the same experiment 10,000 times and we get 6,500 ones (6,500 decays) and 3,500 zeros (3,500 no decays). Now we run the experiment 10,000 times again. Then, from having done similar sets of experiments many times, we know that we will get *close to* 6,500 ones and 3,500 zeros. That is, after many runs of the experiment, the *fraction* of zeros and ones is not just any number between 0 and 1. Instead it will be near .65 for ones and .35 for zeros. That is, there is a probability of .65 of obtaining a one and a probability of .35 of obtaining a zero.

**C. Probability and the wave function.** There is a formula linking a property of the wave function and probability. Suppose we focus on a single nucleus. At time 0, say, the nucleus is not decayed, so its wave function is [not decayed]. But after 5 seconds, its wave function is a combination of [not decayed] and [decayed]. This is written as a sum in quantum mechanics, so we have

[wave fcn after 5 sec]=a(0)[not decayed]+a(1)[decayed]

The a(0) and a(1) are 'coefficients;' their values—actually their values squared—tell 'how much' of the wave function corresponds to not decayed and how much to decayed. Pictorially, $|a(0)|^2$ and $|a(1)|^2$ tell 'how much' of the 'mist' making up the wave function is in the not-decayed aspect and how much is in the decayed aspect.

The link to probability is the following: Suppose

$$|a(0)|^2 = .35, \quad |a(1)|^2 = .65.$$

Then the probability of *perceiving* [not decayed] is .35 and the probability of *perceiving* [decayed] is .65.

**D. The |a(i)|² probability law.** As before, we start out at time 0 in the [not decayed] state. As time progresses, the state will be some combination of [not decayed] and [decayed].

$$[\text{wave function at time t}] = a(0,t)[\text{not decayed}] + a(1,t)[\text{decayed}]$$
$$|a(0,t)|^2 + |a(1,t)|^2 = 1$$

The probabilities for the two states at time t are then

$$\text{probability of not decayed at time t} = |a(0,t)|^2$$
$$\text{probability of decayed at time t} = |a(1,t)|^2$$

This statement generalizes. Suppose at time t, the quantum state of a system is the sum of several possibilities, $\sum a(i)|i\rangle$, with the $i^{th}$ possible 'state,' $|i\rangle$, having coefficient $a(i)$. Then

> **3-1. The |a(i)|² probability law.** If an experiment is run many times, a physical reality with characteristics corresponding to version $i$ will be perceived a fraction $|a(i)|^2$ of the time.

Exactly the same law, but in a different guise, is called the $|\psi(x)|^2$ probability law, or the Born rule [6] (Born first proposed the idea of probability), where $\psi(x)$ is the wave function at position x. (In this form, it is often interpreted as implying that the probability of finding *the particle* at x is $|\psi(x)|^2$. But this is an unwarranted interpretation; see section 7 on the non-existence of particles.)

**E. Conservation of probability.** There are two important points about principle **3-1**. The first is that coefficients $a(i)$ are determined by the Schrödinger equation. The second is that, even though the coefficients may change in time, because of the special nature of the Schrödinger equation, the sum of the coefficients squared is always 1. And so, from principle **3-1**, the sum of the *probabilities* is always 1, exactly as it should be. This is called conservation of probability.

**F. Perception vs. 'actuality.'** Because our world appears to be objective, it is most natural to assume there is a single, actual, objective outcome to an experiment. But one is not allowed, *a priori*, to make that assumption when dealing with the interpretation of quantum mechanics. In fact, as we indicated in section 2, the perceptions predicted by quantum mechanics agree quite well with what we actually perceive without assuming there is just a single version of reality.

**G. Consistency of quantum mechanics proper and the probability law.**
Principle **3-1** cannot be deduced solely from the Schrödinger equation plus the properties of the wave functions. However, suppose we agree that the probability of perceiving outcome $i$ is a function of $|a(i)|^2$ so that $p(i) = f(|a(i)|^2)$. Then there are several ways to show that the only functional form consistent with the mathematics of quantum mechanics is the conventional $f(|a(i)|^2) = |a(i)|^2$. The law $f(|a(i)|^2) = (|a(i)|^2)^2$, for example, would give inconsistent results. I think this observation, that

> **3-2.** The $|a(i)|^2$ probability law is the only functional form consistent with the rest of conventional quantum mechanics.

strongly suggests that the origin of the law must, to a large extent, somehow be *within* conventional quantum mechanics, even though we haven't yet figured out how (see, however, section 11). That is, it doesn't make sense for there to be a probabilistic 'mechanism' that is entirely independent of the laws of quantum mechanics, but *just happens* to be consistent with those laws.

## 4. The Failure of Everett's Many-Worlds Interpretation.

In Everett's many-worlds interpretation [1], which is the bare-bones interpretation of quantum mechanics, it is assumed that (1) only the wave functions, which obey linear Schrödinger equations, exist (no particles or hidden variables); (2) there is no collapse of the wave function; (3) there are no 'sentient beings' outside the laws of quantum mechanics; and (in our version of Everett) (4) there is no *a priori* assumption of a probability law. We call this set of assumptions QMA.

To see the implications of these assumptions, suppose we do an experiment—perhaps a Stern-Gerlach experiment—on an atomic system that has $n$ possible states. Before the experiment, the system has state

$$|\Psi\rangle = \sum_{i=1}^{n} a(i)|i\rangle \tag{4-1}$$

After the measurement, the detector and observer states become *entangled* with the atomic states, and the state is

$$|\Psi\rangle = \sum_{i=1}^{n} a(i)|i\rangle|\det,i\rangle|obs,i\rangle \tag{4-2}$$

This is the state of the universe in the many-worlds interpretation. In it, if you are the observer, there are *n* versions of you! There is an equally valid version of you perceiving each outcome so no version is singled out as *the* you.

To see if this multiple-versions-of-you gives a reasonable interpretation, under the assumption that our current perceptions correspond to those of one of the versions of the observer, we first need to check whether the perceptions of each version violate any of our experiential perceptual rules. From section 2, we see that each version perceives a single version of reality, that two observers never disagree, and that successive measurements are consistent. And from sections 5 and 6, we see that each version will perceive a universe that appears to be made of classical, localized particles. So the many-worlds interpretation, strange as it is, appears to do an excellent job.

**Problems with the probability law.** There is, however, one problem—the $|a(i)|^2$ probability law of principle **3-1**. This law cannot hold within the many-worlds interpretation [7]. We will give a simple form of the argument here, with more detailed arguments given in appendix A5.

We work from Eq. (4-2). First, the only entities that perceive in QMA are the versions of the observer, so all statements about perception must refer to the perceptions of the versions of the observer. Second, one might conjecture that only one version of the observer is 'aware.' But one can show (appendix A6) that all versions are equally aware.

Now each version, $|obs,i\rangle$, of the observer perceives its respective outcome—by looking at the corresponding version, $|det,i\rangle$, of the detector—with 100% probability. There is *nothing probabilistic* in the perception process; there is simply each equally valid version of the observer perceiving its associated outcome. Thus there is no apparent way to introduce probability of perception (principle **3-1**) into QMA.

One might be tempted to argue that the probability comes in the assignment of *my* perceptions to the perceptions of one of the versions; sometimes *my* perceptions correspond to those of version $|obs,1\rangle$, sometimes to those of version $|obs,2\rangle$ and so on, with the probability of assignment to version $|obs,i\rangle$ being $|a(i)|^2$. The problem with this conjecture, however, is that when we say '*my* perceptions,' we are implicitly assuming the existence of a single, unique *me*, separate from the versions. But there is no unique *me* separate from the versions in QMA; if there are *n* versions, there are always *n me*'s. So that interpretation cannot be valid.

**4-1.** The probability law cannot hold within QMA, where all versions of the observer are perceptually equivalent on each run.

Further, we argue in appendix A5 that one version must be singled out as the one corresponding to *my* perceptions.

**4-2.** The probability law implies there must be some mechanism, not within QMA, which *singles out* one version of the observer as the one corresponding to *my* perceptions.

To the best of my knowledge, aside from particles (including hidden variables) and collapse, the only other possible singling out mechanism is the existence of 'sentient beings' whose awareness is outside the laws of quantum mechanics. So principle **4-2** says there must be either be particles (hidden variables) or collapse or sentient beings. Particle interpretations will be considered in sections 5, 6, and 7, hidden variables in section 8, collapse in section 9, and a sentient being interpretation will be given in section 11.

# 5. Group Representation Theory.

In this and the next two sections, we will explore the possibility of particle interpretations of quantum mechanics. We find there is no evidence for particles. In this section, we will show that the properties of mass, energy, momentum, spin and charge, which are attributed to particles in classical physics, can actually be shown to be properties of the wave function of quantum mechanics. The method used is the mathematical discipline of group representation theory, which I will endeavor to explain in a way suitable for non-specialists.

**Physical and mathematical invariance.** Ignoring gravity, we don't expect the results of experiments to depend either upon the orientation (east-west, north-south) or the position of the experimental apparatus. This constitutes *physical* rotational and translational invariance (no variation in the results for changes in orientation or position).

The *equations* governing physical phenomena should reflect this invariance. Suppose we take the Schrödinger equation for the hydrogen atom. Then it should be (and is) independent of how the *xyz* coordinates are oriented (so long as the three axes are perpendicular). So the equation is invariant under any *rotation* of the coordinate system (that is, the form of the equation is the same no matter whether you express it in terms of the original coordinates, xyz, or in terms of the rotated coordinates, x´y´z´). The set of all possible rotations, along with the rules for how two successive rotations give a third, constitute a mathematical entity called the three-dimensional rotation group.

**The hydrogen atom and the invariance chain.** Because the equation for the hydrogen atom wave function is invariant under (the group of) rotations, the solutions to the equation can be classified or *labeled* according to their angular momentum, or spin as it is usually called (see appendix A2). But this is not just a mathematical classification scheme. If a Stern-Gerlach (A2) experiment is done on the atom, one actually gets different, *measurable* results for the different spins. Thus we have this interesting chain—from physical invariance (rotations shouldn't matter in the outcome) to mathematical invariance to a mathematical classification scheme for different wave functions to an actual, physical, measurable property of the wave function.

**The full set of physical invariances.** Physical laws are not just invariant under three-dimensional rotations. They are also invariant under relativistic 'rotations' in the four dimensions of space and time. Further, they are invariant under translations in space and time.

There is also one more kind of invariance, that of the 'internal' symmetry group. Physicists had noticed that there are certain regularities in the properties of the many 'elementary' particles discovered. It has been found that these regularities correspond to invariance under a set of non-space-time rotations—for example rotations in a complex three dimensional space for quarks or perhaps a complex five [8,9] or six dimensional space if electrons and neutrinos are included.

**Mass, energy, momentum, spin and charge.** As in the hydrogen atom case, the physical invariances lead, through the above chain of reasoning, to physical, measurable properties of the solutions to the invariant equations. These properties are mass, energy, momentum and spin for the space-time group [10], and the strong, electromagnetic and weak charges for the internal symmetry group. Thus all the properties—mass, …, charge—that we attributed to particles in classical physics are actually seen to be properties of the quantum mechanical wave function!!

> **5-1.** Linearity and the physical invariance properties—relativistic rotations, translations, internal symmetries—imply the particle-*like* properties of mass, energy, momentum, spin and charge are properties of the wave function.

> **5-2.** Linearity and the invariance properties also imply that the usual conservation laws and laws of addition for energy, momentum, spin and charge hold in quantum mechanics.

One might object that the quantities derived from group representation theory actually refer back to properties of the particles 'associated with' the wave functions. But all the mathematical apparatus used above referred only to the wave functions. There is no reason, at least on this account, to add an extra concept (particles) to the structure of physical existence when that concept never occurred in our deliberations.

**Significance of this result I.**  It is truly astonishing that the existence of the centuries-old *classical* properties of particles, mass, …, charge, is *derived* in quantum mechanics.

The inputs to this derivation are the invariance principles and the linearity of the quantum mechanical equations.  The fact that the consequences of these inputs give the known properties of particles is probably as close as one can come to a proof that invariance and linearity are absolute principles in the description of physical existence.

**Significance of this result II.**  The fact that the classical particle-like properties are properties of the wave functions severely undermines the rationale for assuming the existence of particles.

**Significance of this result III.**  The physical particle-like states of quantum mechanics, in ket notation, are written as $|m,E,p,S,s_z,Q\rangle$ (mass, energy, momentum, spin, z component of spin, and the three charges).  That is, *all the labels on states are group representational quantities*.  Further, the antisymmetry of fermions and the symmetric statistics of bosons are also group representational properties (associated with the permutation group).  This suggests that quantum mechanics as we know it is the *representational form* of an underlying, pre-representational, linear, appropriately invariant theory (see [11]).

# 6. Localization.

There is one particle-like property that is not related to group representation theory, that of localization.  Speaking classically, the carriers of mass and charge seem to be very small, almost point-like, with highly localized effects.  This particle-like localization occurs in an interesting way in quantum mechanical experiments.  To illustrate, suppose we have a single light wave function that goes through a single slit and impinges on a screen covered with film grains.  The wave function spreads out after going through the slit and hits many grains of film.  But surprisingly, a microscopic search will show that only one of the grains will be exposed by the light.  It is *as if* there were a particle of light, a photon, hidden in the wave function, and it is the single grain hit by the particle that is exposed.

As another example, suppose we have a target proton surrounded by a sphere coated with film grains on the inside.  An electron (electron-like wave function) is shot at the proton and the wave function of the electron spreads out in all directions, hitting every grain.  But again, a microscopic examination will show that one and only one grain is exposed.  As in the case of light, it is *as if* a particulate electron embedded in the wave function followed a particular trajectory and hit and exposed only one grain.

**Quantum mechanical explanation.** Surprisingly, this exposure of only one localized grain by a spread-out wave function can be accounted for strictly from the principles of quantum mechanics, without invoking the existence of particles. It depends on the linearity of the theory.

We suppose in the electron scattering example that there is an interaction, represented by $V$ in the equations below, of the electron wave function, $|elec\rangle$, with each of the N grains, $|gr\rangle$. Just before the wave function hits the grains, we can schematically write the state and electron-grain interaction as

$$|elec\rangle [V(gr1)+...+V(grN)]|gr1\rangle...|grN\rangle =$$

$$|elec\rangle V(gr1)|gr1\rangle|gr2\rangle...|grN\rangle +$$
$$|elec\rangle V(gr2)|gr1\rangle|gr2\rangle...|grN\rangle +...+$$
$$|elec\rangle V(grN)|gr1\rangle|gr2\rangle...|grN\rangle = \qquad (6\text{-}1)$$

$$|elec\rangle V(gr1)|gr1\rangle|gr2\rangle...|grN\rangle +$$
$$|gr1\rangle|elec\rangle V(gr2)|gr2\rangle...|grN\rangle +...+$$
$$|gr1\rangle gr2\rangle...|elec\rangle V(grN)\| grN\rangle$$

where we have used allowable rules of re-arrangement of the N terms to arrive at the final form. In the first term of the last three lines, the electron wave function, through the grain 1 interaction term, interacts with and exposes only grain 1. In the second, only grain 2 is exposed, …, and in the last term, only grain N is exposed. Thus the final state is (with the asterisk indicating an exposed grain)

$$|elec\, in\, gr1\rangle|gr1\rangle^*|gr2\rangle...|grN\rangle +$$
$$|gr1\rangle|elec\, in\, gr2\rangle|gr2\rangle^*...|grN\rangle +...+ \qquad (6\text{-}2)$$
$$|gr1\rangle|gr2\rangle...|elec\, in\, grN\rangle|grN\rangle^*$$

> [We have assumed that in each term, the electron—or rather a portion of the electron-like wave function—ends up lodged in the exposed grain. Note that the small portion of the electron-like wave function that lodges in each grain carries the full mass, charge and spin that we associate with an electron—see appendix A7—so that each of the N versions of reality has the correct charge and spin.]

That is, there are N terms and only one grain is exposed in each term. But we know from section 2 that only one term will be perceived. And since each term has just one grain exposed (no matter what the size of the grains), only one *localized* grain will be perceived as exposed, even though the electron wave function hits all the grains! Thus

**6-1.** A spread-out wave function produces localized effects.

It is remarkable that quantum mechanics alone, with no assumption of the existence of particles, can imitate, in our perceptions, the classical idea of a localized (only one localized grain exposed) particle.

**Particle-like trajectories.** In cloud and bubble chambers, a spread-out wave function is found to produce more or less continuous, particle-like trajectories. Reasoning similar to the above, applied several times, shows this also follows from quantum mechanics alone. (The 'continuity' follows because the wave functions of successive nucleation centers are entangled; in each term, the next nucleation center 'nucleates' only if the previous one nucleated.)

# 7. No Evidence for Particles. Wave-Particle Duality.

If one ignores some of the principles of sections 2 through 6, as is often done in interpretations, one way of attempting to explain why we perceive only one of the versions of reality given by quantum mechanics, and why our world appears particle-*like*, is to suppose that the wave function is not the physical reality we perceive. Instead there is an actual, unique physical existence, made up perhaps of particles, which rides along within just one of the versions, and it is that physical reality that we perceive. This, in fact, is the view of the majority of scientists and non-scientists alike. If you look in a typical modern physics text, you will find analyses of experiments—the Compton and photoelectric effects, for example—which reputedly prove particles are necessary for understanding physical existence.

But the problem with these arguments is that they do not take into account *all* the properties—well-known and not so well-known—of the wave function. If these are taken into account, then one can show that all the particle-*like* properties can be explained by properties of the wave function alone.

**A. Wave-particle duality.** It is useful to approach this from the perspective of wave-particle duality. The classically wave-like and classically particle-like properties of matter are:

**Wave-like properties.**
Diffraction (single slit)
Interference (double slit)
Properties of a fraction of a wave.

**Particle-like properties.**
Mass, energy, momentum, spin.
Charge.
Localization and particle-like trajectories.

**B. Wave-like properties.** The first two wave-like properties follow for the wave function from the linearity of the quantum equations of motion. The third property, however, is interesting because it differs for classical waves and the wave function. In a classical wave, a water wave or sound wave, for example, a small part of the wave carries only a correspondingly small part of the energy and momentum of the wave. But it is shown in appendix A7 that

> **7-1.** A small part of a wave function carries the *full* mass, spin, charge, energy and momentum.

What about light waves, which correspond both to classical waves and to photon-like wave functions? A portion of a classical light wave, say a beam of light, is composed of many millions of photon-like wave functions. And so it is indeed true that a portion of the beam carries a corresponding portion of the energy and momentum because it carries a portion of the 'photons'. But for a single photon wave function, a small portion of it carries the *full* energy and momentum. (Energy and momentum add across products of wave functions in quantum mechanics, but not across sums.)

**C. Particle-like properties.** We have already shown in sections 5 and 6 that the classical particle-*like* properties are actually properties of the wave functions.

**D. Compton scattering and the photoelectric effect.** Perhaps the most convincing 'evidence' for the existence of particles comes from these two experiments, which both involve the interaction between light and electrons. In it, if one assumes there are particulate electrons and photons that carry energy and momentum, and if one employs conservation of energy and momentum, then one can explain the experimental results. But we have seen that energy and momentum are properties of the wave functions, and that the conservation laws also hold in quantum mechanics. These, combined with the third property of waves for wave functions (a portion of the wave carries the full energy and momentum), are sufficient to derive the Compton and photoelectric results without assuming the existence of particles (either photons or electrons). Thus, because quantum mechanics alone, without the presumption of particles, can explain Compton scattering and the photoelectric effect, these two experimental results do not provide evidence for the existence of particles.

The same conclusion holds for all other experiments that allegedly give evidence for the existence of particles. Thus,

**7-2.** There is no evidence for the existence of particles.

And since there is no evidence, there is no reason to assume particles exist.

      **E. Nomenclature.** Even though there is no evidence for particles, it is not necessary to give up the very useful names. So we can agree that an 'electron' refers to an electron-like wave function, having mass $m_e$, spin ½, charge, $-e$; and a 'photon' refers to a photon-like wave function having mass 0, spin 1, and charge 0. But we must keep in mind that these names do not (if there is no collapse) refer to entities that objectively exist in a unique state; there can be several versions of them simultaneously existing in different states.

      **F. Related experiments.** There are several experiments that, if one assumes particles exist, give perplexing results. Bell-like experiments (appendix A8) [12,13] on non-locality seem to show that particles interact instantaneously at a distance. And Wheeler's delayed-choice experiment (A9) [14,15] and the quantum eraser experiment (A10) [16,17] seem to imply that causality can sometimes work *backwards* in time. But if one assumes there are no particles, the very peculiar inferences about the nature of reality simply evaporate, as is shown in appendices A9 and A10. In a no-particle world, there is (to the best of my knowledge) no evidence for backwards causality. And while there are non-local effects in quantum mechanics, built into the theory by entanglement, there is no faster-than-light signaling required to explain the results; it is just quantum mechanics as usual.

# 8. Hidden Variable Interpretations.

      It has been conjectured that even though one cannot show there are particles, there may still be 'hidden' (not detectable by experiment) variables that determine the unique outcome of an experiment, the outcome we perceive. At a given time, there is only one, unique set of these hidden variables (as opposed to the many versions of reality of the wave function), so they are objective.
      The first observation is that

**8-1.** There is no experimental evidence for hidden variables.

Second, there is no definitive proof that there cannot be a hidden variable theory underlying quantum mechanics. And third, there is no acceptable hidden variable theory at this time, and one encounters severe difficulties in attempting to construct a satisfactory underlying theory.

**Bohm's model.** The most nearly successful hidden variable theory is that of Bohm [18,19]. Although it falls short, there are several reasons for presenting it. First, it *is* successful in achieving its goal if one ignores certain shortcomings. Second, it is pretty much the only even minimally acceptable hidden variable model. Third, any acceptable hidden variable theory would presumably have to reduce to Bohm's model in many situations. And fourth, it serves as a testing ground for conjectured 'no-go' theorems—arguments which purport to prove that hidden variables are impossible.

In this model, Bohm derived a set of trajectories through space from the Schrödinger equation (the velocity of the trajectory through point x is essentially the gradient of the phase angle, $\varphi(x)$, of the wave function). He then assumed that for each 'particle-like' wave function, a 'particle' was put on one of the trajectories. This particle then followed a very complex trajectory. The hidden variables in this case are the position and velocity of the particle.

To explain the success of the theory, we will use the example of a Stern-Gerlach experiment [20] (see appendix A2) on a spin 1 particle (particle-like wave function). The z-component of spin of this wave function can take on the three values $1, 0, -1$. When this wave function is shot through a magnetic field, the three parts of the wave function each follow separate trajectories, so the wave function, schematically, is

$$\Psi = a(1)\psi_1(x) + a(0)\psi_0(x) + a(-1)\psi_{-1}(x) \qquad (8\text{-}1)$$

where the different wave functions $\psi_1(x), \psi_0(x), \psi_{-1}(x),$ correspond to the different trajectories. For each run of the experiment, the probability law tells us that the probability of the particle following trajectory *i* is $|a(i)|^2$.

We now suppose the spin 1 particle consists of two spin 0 particles in a bound state (like the electron and proton in a hydrogen atom, ignoring the spin ½ ) and apply the Bohm model to this situation. Before the particle reaches the magnetic field, the two particles bound together will follow trajectories that *move very rapidly* through the three possible spin 1 states. And then at the magnetic field, the trajectory of the bound state will follow just one of the three possible paths. The success of Bohm's model is that it predicts that path *i* will be followed a fraction $|a(i)|^2$ of the time, so that the probability law is satisfied.

Now, for the problems with the model. First, it is not relativistic, and because of the way in which time is used, it is difficult to generalize it to a relativistic formulation. Second, there are as-yet-unsolved problems in handling the creation and annihilation of particles. Third, one must assume a specialized initial density of trajectories in order to obtain the probability law, and there is no clear reason why nature should choose that density.

Fourth, it is an arbitrary feature of the model that a particle is put on just one of the trajectories. There is nothing in the mathematics that prevents there being two particles, on two different trajectories, associated with a 'single particle' wave function [21]. And it is extremely difficult to reformulate the Bohm model with a 'source' equation which forces there to be one and only one particle associated with a single particle wave function [21].

Finally, there is the problem of perception. The wave function does not collapse and so there is a valid quantum state of the observer's brain corresponding to each outcome. There is nothing in the mathematics of the theory which says that only the version of the brain associated with the particle is 'consciously aware.' To put it conversely, an acceptable no-collapse hidden variable theory must give a convincing reason for why the *non*-particled quantum versions of the observer are not aware. Such a reason is not given in the Bohm model, and I don't believe it can be given in any hidden variable model. Thus we have

> **8-2\*.** It appears to be very difficult, perhaps impossible, to explain in *any* non-collapse hidden variable model why the quantum versions of the brain *not* associated with the hidden variables cannot be consciously aware.

with this principle being less certain than the others, hence the asterisk. One cannot simply say "Well, that's just the way Nature works" because, in 'being-less' hidden variable schemes, there is no arbiter, no outside intelligence that could make the choice between particled and not-particled quantum versions. And there is nothing that is aware of only particles or their state (and the particles themselves are not assumed to be aware). There is not even a current understanding of what conscious awareness *is* to base one's argument on.

If one takes a different tack and conjectures that the hidden variables change the structure of the 'particled' quantum version of the brain in some subtle way, that is a whole different kind of hidden variable theory. It runs up against the successes of no-hidden-variable quantum mechanics.

**The Kochen-Specker reasoning.** Kochen and Specker [22], and later Conway and Kochen [23], used a most elegant argument in an attempt to show that it is not possible to have an underlying hidden variable theory. The primary characteristic of a hidden variable theory is that,

> **8A.** Once the hidden variables (whatever they are) are set, the outcome of *any* experiment is determined.

Kochen and Specker then showed that for certain sets of experiments, this leads to a contradiction. That is, they claim it is logically impossible to satisfy principle **8A** in every situation! If this is correct and free of hidden assumptions, then it would constitute a welcome proof (certainty is sometimes hard to come by in the interpretation of quantum mechanics) that there could be no hidden variable theory.

The problem is that there is an apparent counter-example, the Bohm model, to the K-S-C no-go theorem for hidden variables. For this reason, I believe that K-S-C did indeed make a hidden assumption that would appear to invalidate their result unless they can work around it. The argument is given in detail in appendix A11, but the gist of it is that they implicitly assume the hidden variables are the *only* quantities that determine the outcome of a measurement, and that is not correct.

Thus, as far as I know, even though the construction of a satisfactory underlying hidden variable theory is fraught with difficulties, and is *probably* not possible, there is no *proof* (under acceptable assumptions) that it cannot be done. (See appendix A8 for a non-local restriction of sorts imposed on hidden variable theories by Bell-like experiments.)

# 9. Collapse Interpretations.

Another way of attempting to explain why we perceive only one version of reality is to suppose that the wave function somehow *collapses* down to just one version [4]. In Schrödinger's cat, for example, the dead cat version of reality might collapse to zero—that is, it would simply go out of existence—leaving just the live cat version to be perceived.

There have been a number of searches for experimental evidence of collapse, including interference effects for large molecules, SQUID experiments, and constraints set by nuclear processes [24-26]. But none of them has found any evidence whatsoever for collapse.

> **9-1.** In spite of a number of attempts, no experimental evidence for collapse has been found.

(Note also section 3F. The simple fact that we perceive only one version of reality does not constitute experimental evidence that the others have collapsed to zero.)

**Mathematical theories of collapse.** The primary (and just about the only) *mathematical* theory of collapse is the one proposed by Ghirardi, Rimini, Weber, [27]and Pearle [28]. A random force is introduced into the time evolution of the wave function in such a way that, after a short period of time, a microsecond or less for systems with many particles, there is a collapse of the wave function down to just one version (without affecting the 'shape' of the wave function). The coefficients of all the other versions effectively shrink to zero, so those versions simply go out of existence. This idea is beautifully implemented in the Pearle model. But even though the mathematics is elegant, there are difficulties with the physical implications of the scheme [26]:
- •There must be *instantaneous* coordination of billions of random events located far from each other (and for that reason, one is currently not able to make the theory relativistic).
- •The coordination extends *across versions*, which is strictly forbidden in conventional quantum mechanics, because versions don't communicate.

- The random event 'chosen' at time *t* depends on the *results* of the choice at time *t*, (rather than at time $t - dt$ as is usual in a physically applicable differential equation) which seems physically awkward.
- The linearity of quantum mechanics must be abandoned (appendix A12), which runs counter to the group representational analysis in section 4, as well as to all the successes of quantum mechanics.
- The specific Pearle method of collapse, using particle number, doesn't work in all situations.
- Finally, the specific and very specialized form *assumed* (not derived) for the randomizing Hamiltonian *just happens* to give the $|a(i)|^2$ probability law which agrees with the rest of quantum mechanics (see principle **3-2** and the comment at the end of section 3).

Because there are no other serious candidates for a mathematical theory of collapse, and also because some or all of these problems would be expected to occur for any model, there is currently no reason to suppose that mathematically implemented collapse is the solution to the problem of perception of only one version of reality.

> [Actually the problem is not the perception of only one version—see section 2; the problem is why the probability law holds. The GRW-P model does indeed give the probability law, but it has the problems just described.]

**Collapse by conscious perception.** To summarize the reasoning of section 7, we have seen that there is no evidence for the existence of anything besides the wave function. We have also seen that there is no experimental or theoretical evidence for a *mathematical* theory of collapse to just one version. And yet we know from section 4 that *one version must be singled out for perception*, by some means, if the probability law is to hold. One possible singling out mechanism, the one addressed here, is that there is indeed collapse, but it is initiated by *conscious perception*, rather than by a mathematical process based on random variables [4,29].

However, because an instantaneous collapse, $\sum a_i |i\rangle \to |j\rangle$, suspends the usual Schrödinger equation mathematics for an instant, this option has the disadvantage that the conscious perception of a human being—a slippery concept, subject to wandering attention, impaired perception and so on—disrupts the mathematics. Further, the collapse changes the wave function in such a way that even though the Schrödinger equation has been suspended, the wave function is still a solution to the equation when it comes out the other side of the collapse. In addition this instantaneous process adheres to the probability law, collapsing to state $|j\rangle$ on a fraction $|a(j)|^2$ of the runs. This creates difficulties which are discussed in appendix A13. This consciousness-triggered intervention in the mathematics, tailored to the situation with no apparent way of verifying the assumptions, and with *no explanation* for why the 'collapsing agent' adheres to the probability law (see principle **3-2**), does not seem to me to be a likely way for Nature to proceed.

Thus we have the following principle:

**9-2.** There is currently no reason to be at all optimistic that there is a theory of collapse which augments standard quantum mechanics.

**Decoherence and Consistent Histories 'Interpretations.'** The ideas of decoherence [30,31] were not meant as an interpretation but they are sometimes construed that way. In decoherence, the environment often causes a system to collapse to just one version. But the environment cannot do that in general. Suppose we do a Stern-Gerlach experiment on a spin ½ silver atom. As soon as the wave function of the silver atom separates into two non-overlapping parts (as it goes through the magnetic field), the universe is divided into two separate, non-interacting versions (see appendix A3), each of which keeps its same norm (same size, same coefficient). Nothing in the detectors or the environment in general can change the norms of the two parts once they separate. Thus there can be no collapse. There is therefore no singling out mechanism, and so the physical process of decoherence cannot provide a satisfactory interpretation of quantum mechanics. The same is true for Consistent Histories [32]; it provides no singling out mechanism in the above example.

**Penrose's gravitational collapse.** Penrose [33] proposed the following (in connection with a theory of consciousness): Suppose we again do the Stern-Gerlach experiment so there are two versions of the detecting apparatus and the observer. Since the arrangement of atoms is slightly different in the two versions, the definition of space and time—which depends on the distribution of matter—will be slightly different in the two versions. Penrose supposed that difference caused a 'tear' in the fabric of space-time. This tear would lead to a force that would pull the two versions together, thus causing a collapse to just one of them. But this inter-version force violates the principle that versions are in separate universes which cannot influence one another. So unless Penrose can show why this separate-universe principle doesn't hold, I don't believe his interpretation can be valid.

# 10. Implications of the Principles.

The interpretation we will consider in section 11 makes use of a "Mind" outside the laws of the physical universe. So, to motivate this radical step, and also to bring some order to the interpretive problem, we will review our results so far. The question to be answered is: What underlying conceptual picture of reality can best link our perceptions with the mathematics of quantum mechanics?

We start with the principle

### 10-1. Physical reality consists of the wave function alone.

To me, the evidence is overwhelming that physical reality consists of the wave function alone. The particle-like properties of mass, energy, momentum, spin, charge and localization all follow from the principles of quantum mechanics alone. And principle **2-2**, that one version of the wave function always agrees exactly with our perceptions, makes it a virtual certainty.

This implies that interpretations which postulate a real classical world are not likely to prove fruitful.

Further, there are interpretations in which it is assumed that quantum mechanics gives only statistical information, or information about how we get from one state to another, but they do not say anything about the nature of the matter itself. I believe such interpretations are unnecessarily conservative. Further, they do not give sufficient weight to the non-statistical or 'absolute' facts that quantum mechanics implies matter has mass, charge, quantized spin, and is perceived as localized; and that it gives accurate energy levels for composite systems.

So we will suppose that the only interpretations worth considering at this point are those in which physical reality consists of the wave function alone. This is not a done deal, because there is no *proof* that hidden variables do not exist, or that there is no collapse. But in view of our review of the current state of physics, it seems like a quite reasonable assumption.

**Can we use physics to infer the nature of reality?** There is a point of view which says that, because we don't directly perceive anything other than our macroscopic world, we have no business inferring anything about the true nature of reality from quantum mechanics. I just don't buy this no-inference-allowed argument. In fact, I think it works better in reverse. What we are *directly* aware of corresponds solely to our *neural representation* of reality. But mathematical physics, the distillation of the results of millions of experiments (with experiments just being disciplined, organized perceptions), gives a much more detailed, in-depth, and unified picture of the physical world than our sense-based view. Thus the picture of the physical world that mathematical physics gives us—that we live in a world made of wave functions (see also reference [11] on an underlying theory)—is more likely to correspond to a scheme which is closer to the 'actual nature' of reality than the everyday picture given by our neural representation.

A second major principle is

### 10-2. It is quite reasonable to assume there is no collapse.

This follows from our observation that there is no experimental evidence for collapse, from the difficulties encountered in attempting to construct a valid mathematical theory of collapse, and from the questionable suggestion that collapse is based on conscious perception.

Third,

**10-3. Even if there is no collapse and there are no hidden variables, the probability law implies there must be some 'mechanism' which singles out one version for perception on each run.**

And fourth,

**10-4. Since there is no singling out mechanism in basic quantum mechanics** (the QMA of section 4)**, the mechanism for singling out the perceived version of reality must be outside the laws of basic quantum mechanics.**

Given 10-1 and 10-2, I can think of only one way to single out one version. It is to suppose there is a "Mind," outside the laws of quantum mechanics, which is the source of the awareness necessary for perception. One version of this, the Mind-MIND interpretation, is given in the next section.

# 11. The Mind-MIND Interpretation.

The arguments given in sections 5 through 8 show we can be quite (although not absolutely) certain there are no particles or hidden variables. And the arguments in section 9 indicate we can be fairly certain there is no mathematical or consciousness-triggered collapse. We thus arrive at the following set of 'idealized' requirements on an interpretation.

**11-1.** There are no particles or hidden variables.

**11-2.** There is no collapse.

**11-3.** One version must be singled out as the one corresponding to my perceptions on each run (from section 4).

How do we proceed from here? To see, suppose we go back to principle **2-2**—that that our perceptions always correspond exactly to those of one and only one version of the observer. It is *as if* there is a "Mind," outside the laws of quantum mechanics, that is, in some sense, concentrating on and perceiving one quantum version of physical reality. This supposition forms the basis for the Mind-MIND interpretation.

## A. Basics of the Mind-MIND Interpretation

**1. The non-physical Mind.** Associated with each individual person is an individual Mind that is not subject to the mathematical laws of quantum mechanics. In particular, the Mind has no wave function associated with it. We use a capital M to distinguish this Mind from the usual usage of the word mind. Because quantum mechanics describes the physical world so well, we will *define* anything outside its laws to be 'non-physical.' So with this definition, the Mind is non-physical.

**2. The Mind perceives only the brain-body.** The individual Mind perceives only the wave function of the individual brain (brain-body); it does not directly perceive the quantum state of the external world. This is consistent with the fact that we are not directly aware of the external world; we are only aware of the neural state of our brain. The process of perception of the wave function by the Mind is not understood.

**3. The Mind picks out one version of the wave function.** The individual Mind concentrates on one version of the wave function of the brain-body, and it is the concentrated-upon version that enters our conventional awareness.

**4. No collapse.** The Mind does not collapse the wave function or interfere with the mathematics in any way.
> A primary objection to dualism—a non-physical Mind separate from a physical brain-body—is that the non-physical aspect must exert a force or otherwise have some effect on the physical world. The Mind scheme circumvents this objection because the non-physical aspect only *perceives*; it does not affect the physical world (which is made up of wave functions) in any way.

# B. Agreement among Observers.
# The Overarching MIND.

The model as it has been given so far leaves two important questions unanswered—why observers agree on what they perceive, and why the probability law holds. To make the first question specific, consider again the Schrödinger's cat experiment and suppose we have two observers. Then according to the rules of quantum mechanics, the wave function is

[cat alive]
[obs 1's brain state corresponds to cat alive]
[obs 2's brain state corresponds to cat alive]
—*and*—
[cat dead]
[obs 1's brain state corresponds to cat dead]
[obs 2's brain state corresponds to cat dead]

Suppose observer 1's Mind focuses on the version of the associated brain corresponding to cat alive so that observer 1 perceives, in the everyday sense, a live cat. We know from everyday experience that observer 1 and observer 2 (and the cat) must be in agreement. And we know from property **2-4** that two observers can never disagree. But still, how do we guarantee in our Mind model, that observer 2's Mind is also focused on the 'alive' version of its brain? There is a way to bring about agreement but it is bound to make scientists even more skeptical of this proposal because it is far outside the realm of contemporary science.

> **5. The overarching MIND.** Instead of each individual Mind being separate from all others, each Mind is a fragment or facet of a single overarching MIND. Each individual Mind is that aspect of MIND that is responsible for perceiving the state of the associated individual physical brain. Concentration of perception on a particular version of the wave function by one individual Mind is then presumed to set the concentration of perception on that same version by the overarching MIND. And that in turn sets the perception of the same version by all the other individual Minds. Neither the MIND nor the individual Minds alter the wave function in any way.

So to obtain conscious agreement among observers in this scheme, we are forced to substitute the MIND assumption for the conventional particle or collapse assumptions that give a single-version physical world. The scientist will say there is no evidence to justify such an outrageously non-scientific assumption. But the counter-argument is that there is no evidence to justify the particle or collapse schemes either. And if there is no collapse and there are no particles—the option we are exploring here—we are apparently *forced* to a "Mind" interpretation. Agreement among observers then forces us to the Mind-MIND scheme.

**Freedom of choice.** Freedom of choice enters the Mind-MIND interpretation in the sense that the individual Mind can *choose* to perceive any possible *internal* quantum state—states in which the branches are branches of the *brain* wave function *alone*; that is, the freedom applies to thoughts and preparations for muscular actions. (See appendix A14 for the argument that the brain has many simultaneously existing internal quantum states.) But this freedom of choice of the individual Mind does not apply to *external* events, where the branching includes objects besides the brain. We cannot *choose* to perceive a live Schrödinger's cat instead of a dead one, nor can we choose to perceive a decayed nucleus instead of an undecayed one.

## C. The Probability Law.

There is a problem with the concept of probability if we assume no particles and no collapse. Probability as it is traditionally understood, say in the dice-rolling example, refers to the probability of a specific, actual, 'single-version' event; there is an actual die that has a specific reading after each roll. But in quantum mechanics, the wave function does not give a specific, actual, single-version event; instead, it gives several versions of reality (equivalent to all six readings of the die occurring at the same time), each potentially corresponding to an 'actual, perceived' event. So the direct use of classical probability is not appropriate in the Mind interpretation, where all versions exist forever. (See also section 4.)

In spite of this problem, there is a way to salvage the probability law in the Mind-MIND scheme. To do this, we make the following (relatively weak) assumption:

> **6. Probability.** The overarching MIND is 'much more likely' to perceive a version of reality that has a much larger norm than other versions. The MIND doesn't have to follow the usual $|a(i)|^2$ probability law or any stable probability law at all; it just has to be much more likely to perceive those versions with a relatively large norm.

To see how this leads to the $|a(i)|^2$ probability law, suppose we do a spin ½ Stern-Gerlach experiment (appendix A2) $N$ times, with $N$ large. The amplitude for spin ($+$ ½) is $a_1$ and that for spin ($-$ ½) is $a_2$. Then if the observer observes only the end result—and not the intermediate ones—the amplitude squared for perceiving $m$ ($+$ ½) spins and $N - m$ ($-$ ½) spins is

$$|A(m,N)|^2 = \frac{N!}{m!(N-m)!}|a_1|^{2m}|a_2|^{2N-2m}, \quad |a_1|^2 + |a_2|^2 = 1$$

A simple calculation using Stirling's approximation for the factorials shows that the amplitude squared has a sharp maximum, as a function of $m$, at $m=N|a_1|^2$, so the amplitude squared for $m$ near the maximum is much larger than the amplitude squared for $m$ not near the maximum. Thus according to property 6, the *perceived* result will have $m$ near the maximum at $N|a_1|^2$. (See also appendix A15.)

> It is worth emphasizing two points here.
> •The probability law in the Mind-MIND interpretation only pertains to a large number of repetitions of the experiment (which, of course, is pretty much true in classical probability theory also).
> •The probability law for long runs holds in the Mind-MIND interpretation only if individual results are *not* observed (which is quite different from classical probability theory).

The second point is most interesting because it implies this view of the probability law can be experimentally tested. Suppose the Stern-Gerlach (or some other) experiment is carried out in such a way that the observer perceives the outcome of *every* run. Then if the above holds, it is possible, indeed likely, that after many runs of the experiment, the average value of $m/N$ could be relatively far from $|a_1|^2$. For further comments on this possibility, see appendix A15.

**The MIND must perceive the magnitude of the coefficients.** There is one more observation here. In the last problem considered in appendix A13, we noted that, to obtain information on the coefficients, the collapsing 'mechanism' must 'perceive' the whole wave function, not just that of the brain. In our scheme, this means that for events external to the brain—Schrödinger's cat, nuclear decay—it must be the overarching MIND that perceives the *whole* wave function and then decides which version to concentrate on. (Remember there is no collapse here, so no interference effects inadvertently get wiped out.) But the individual Mind—a restricted aspect of the MIND—can still perceive just the *brain* wave function and then decide on which *internal* state of the brain (that is, which thought) to concentrate on.

## D. Related Thoughts.

**The many-minds interpretation:** In the many-minds interpretation of Albert and Loewer [34,35], there is a *continuous infinity* of minds associated with *each* observer. Each of these minds chooses to perceive one or the other branch at random, according to the $|a(i)|^2$ probability law. And so, assuming our current perceptions correspond to those of one of the continuous infinity of minds, we would *seem* to arrive at the probability law—our perceptions are more likely to correspond to a state with large $|a(i)|^2$ because more minds have chosen that state.

But that is not correct. Suppose we have a two-state system with $|a(1)|^2 = .2, |a(2)|^2 = .8$. There is a *continuous infinity* of minds that perceive each outcome. But even though four minds choose state 2 for every one that chooses state 1, the *same number* of minds choose each. Why? Because of the way infinities work; four times infinity is still just infinity. Or to say it another way, there are the same number, $\aleph_1$, of points on a line from 0 to .2 as there are on a line from .2 to 1. Thus, because the 'same number' of minds perceives each possibility, the many-minds interpretation cannot account for the probability law.

**Mindless hulks.** One of the main reasons Albert and Loewer proposed their complex many-minds scheme was to avoid mindless hulks—versions of the observer with no associated Mind. But I don't see them as a problem. The 'enlivening' principle in a being is the non-physical Mind. When the Mind does not focus on a particular version of the observer, that version reverts to the same status as any other mindless object, such as a detector; it just evolves according to quantum mechanical laws. The same is true, of course, for the "Minded" version. But the Mind associated with the "Minded" version keeps making *decisions* about which branch constitutes 'reality,' and that is what makes it alive. The Mind is "the fire in the equations ." (Actually it is the fire outside the equations.) Note that we have no sentient beings (the version 'inhabited' by the Mind) talking to mindless hulks (the versions not inhabited) in the Mind-MIND interpretation because each individual Mind and the overarching MIND all concentrate on the same version.

**Panpsychism.** One potential solution to the problem of consciousness is to suppose that matter itself contains or carries intelligence and awareness. But if physical reality consists of the wave function alone, with no collapse, this solution runs into difficulties. Panpsychism, under these conditions, would seem to require that *each version* of a particle wave function has its own intelligence. Further, the intelligences of the different versions would have to cooperate, in accord with the probability law, to somehow *single out* just one version (as required by the arguments of section 4) as the perceived version. This seems awkward.

**Quantum Jumps.** ("If we have to go on with these damned quantum jumps, then I'm sorry that I ever got involved." Schrödinger.) Suppose one does a spin ½ Stern-Gerlach experiment (appendix A2). Then it *appears* that the state jumps from a quantum state in which the wave function is in both states to a quantum state in which the wave function is in only one state, say the + ½ state. But that is not true in the Mind-Mind interpretation; the quantum state remains a linear combination (no collapse) so there is no jump. The only thing that is relevant is our perception of the quantum state; we *perceive* it in one state or another.

# 12. Summary.

In sections 2 through 9, we have given 21 principles that need to be taken into account in constructing a valid interpretive scheme for relating the mathematics of quantum mechanics to our perceptions. And we have shown how these relate to various interpretations. Here, we give the most important points in tabular form. The left column represents basic principles and the right column, consequences.

| | Each version of reality is in a separate universe. |
|---|---|
| Linearity. | There is a version of the observer in each potential version of reality. |
| Many versions of reality. | We perceive only one version of reality. |
| | One version of reality always matches our perceptions. |

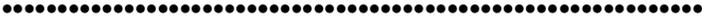

| One quantum version always matches our perceptions. | |
|---|---|
| 'Separate universes' implies localization. | |
| Group representation theory implies mass, charge, etc. belong to the wave function. | No evidence for **particles**. |
| Small parts of the wave function carry the full energy and momentum. | |

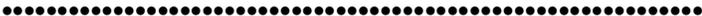

| Linearity, unreasonable interactions, unfruitful experimental search. | No evidence for **collapse**. |
|---|---|

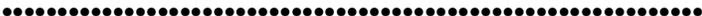

| The probability law. | The **many-worlds** interpretation cannot accommodate probability so it is not valid. |
|---|---|

Now, given the principles and the current state of physics, we list our 'most reasonable' constraints on an interpretation.  The Mind-MIND interpretation is one way to satisfy all these constraints.  In that interpretation, each individual has an associated Mind, outside the laws of quantum mechanics, which perceives just one version of the individual's brain wave function.  And each individual Mind is an aspect of a single over-arching MIND.  This interpretation has the bonus that, with one additional weak assumption, the probability law is automatically satisfied, in accord with principle **3-2**.

One quantum version always
matches our perceptions.

No particles or
hidden variables;
only the wave function.

No collapse, mathematical or
perceptual, so the Schrödinger
equation always holds.

One version is singled out
(required by the probability law).

Communicating versions of two
observers must be on an equal par.

**The Mind-MIND interpretation.**

Probability is a *consequence*, not an input.

# Appendices.

## A1. The Polarization of Photons.

A photon is the name given to a 'piece' of light. In this treatment, where it is assumed (see section 7) there are no particles, it refers to a wave function with mass 0, spin 1, and charge 0. Each photon (wave function) travels at the speed of light. If it has wavelength $\lambda$, its energy is $hc/\lambda$ and its momentum is $h/\lambda$ where $c$ is the speed of light and $h$ is Planck's constant.

Each photon has two possible states of *polarization*, which is analogous to spin or angular momentum. This can be visualized in the following way: A photon can be thought of (at least classically) as consisting of oscillating electric and magnetic fields. If a photon is traveling in the z direction, its electric field can either oscillate in the x direction, so the photon is represented by $|x\rangle$, or its electric field can be oscillating in the y direction, so it is represented by $|y\rangle$. These are the two states of polarization.

It can also be polarized so its electric field oscillates in any direction, say at an angle $\theta$ to the x axis. In that case, the photon state can be represented as a (linear) combination of the $|x\rangle$ and $|y\rangle$,

$$|ph,\theta\rangle = \cos\theta |x\rangle + \sin\theta |y\rangle \qquad (A1-1)$$

There are certain materials (such as Polaroid) that act as filters in that they let through only that part of the photon polarized in a certain direction. In terms of the wave function, if we have a polarizer $[P,x]$ that lets through only $|x\rangle$ photons, then

$$|ph,\theta\rangle[P,x] = \cos\theta |x\rangle \qquad (A1-2)$$

What this translates into experimentally is the following: Suppose we have a beam of light that contains N photons in state A1-1 crossing a given plane per sec, each moving at the speed of light. Then a 100% efficient detector put in the beam will record N 'hits' per second. But if we put an *x* polarizer in the beam, the detector will record only $N\cos^2\theta$ hits per second. And if we put in a *y* polarizer, the detector will record only $N\sin^2\theta$ hits per second.

Thus the polarization state of Eq. (A1-1) acts *as if* it were composed of both a photon polarized in the *x* direction *and* a photon polarized in the *y* direction. But if we measure, we find an $|x\rangle$ photon on a fraction $\cos^2\theta$ of the measurements and a $|y\rangle$ photon on a fraction $\sin^2\theta$ of the measurements. Not so easy to understand!

There are crystals that can separate the photon into its *x* and *y* polarization parts, so that the *x* polarized part travels on one path and the *y* polarized part travels on another path. One can put a detector in each path, and one can have an observer that looks at each detector. If we send a single photon, in state (A1-1), through this apparatus, the wave function is still a sum:

$$|state\rangle = \cos\theta \, |x\rangle \, |Det\, x, yes\rangle \, |Det\, y, no\rangle \, |Obs, x\, yes; y\, no\rangle + $$
$$\sin\theta \, |x\rangle \, |Det\, x, no\rangle \, |Det\, y, yes\rangle \, |Obs, x\, no; y\, yes\rangle \quad \text{(A1-3)}$$

But now the whole 'universe,' photon, detector and observer, has split into two states. In particular, if you are the observer, there are two states of you! On a fraction $\cos^2\theta$ of the runs, the "*x*, yes; *y*, no" version will correspond to your perceptions and on a fraction $\sin^2\theta$ of the runs the "*x*, no; *y*, yes" version will correspond to your perceptions. On a given run, quantum mechanics does not say which version will correspond to your perceptions. So this experiment and the state of Eq. (A1-3) give another example of quantum mechanics presenting us with more than one version of reality—one version of you perceiving polarization *x and, at the same time,* another version of you perceiving polarization *y*.

## A2. Spin and the Stern-Gerlach Experiment.

Discrete values of angular momentum, or spin as it is called on the atomic level, are one of the hallmarks of quantum mechanics. The Stern-Gerlach experiment, used to measure spin, is one of the most-used examples for illustrating ideas in quantum mechanics. So it seems useful to gather all the relevant ideas in one place.

**The invariance chain of reasoning.** As we noted in section 5, rotations of the whole apparatus do not affect the outcome of experiments. This implies the equations must have the same form in all rotated coordinate systems. This form invariance in turn implies that solutions of the equation can be labeled by spin or angular momentum. And we further find that the spin, even though it is initially just a label on a solution, can actually be measured, so that the labels have physical meaning. That is, we have this chain of reasoning—from physical invariance (rotations shouldn't matter in the outcome) to mathematical invariance to a mathematical classification scheme for different wave functions to an actual, physical, measurable property of the wave function.

**Classical and quantum angular momentum.** Angular momentum was defined in classical physics centuries ago; it is the 'momentum around the center' (momentum times the radius) a 'particle' has when traveling in a circle. This could take on any value so that classically, angular momentum is a continuous variable.

But that is not true in quantum mechanics. Instead, when measured along a certain direction, angular momentum (spin) can only take on certain discrete values. For a spin 1 wave function, the allowed values (in units of the Planck constant $\hbar$) are 1,0,-1. For a spin 2 wave function, they are 2,1,0,-1,-2, and so on. This *quantization* (only a discrete number of values allowed, instead of a continuum) of angular momentum is one of the primary characteristics of quantum mechanics.

**The Stern-Gerlach experiment.** Angular momentum or spin is measured on an atomic scale by the Stern-Gerlach experiment. A particle (particle-like wave function) is shot into a magnetic field and the field exerts different forces on the parts of the wave functions having different values of spin. Because of the differing forces, the parts of the wave function with different spin exit the magnetic field traveling in slightly different directions. Suppose, for example, that a spin 1 particle is shot into the magnetic field. Then three different streams of particles will come out of the apparatus; the +1 spin might be traveling slightly upward, the spin 0 would travel straight through, and the -1 spin would be travelling slightly downward. Three different detectors could be put in the three paths to determine which path the particle took (this 'particle' and 'which path' language is convenient but misleading).

**Many versions of reality.** So let's suppose the wave function for the particle before it reaches the magnetic field is a linear combination of the three possible spins,

$$|before\rangle = a(+1)|+1\rangle \psi(x) + a(0)|0\rangle \psi(x) + a(-1)|-1\rangle \psi(x) \qquad \text{(A2-1)}$$

where the kets, $|\ \rangle$ indicate the spin part of the wave function and the $\psi(x)$ represents the spatial part of the wave function (the same for all three here). After the particle goes through the magnetic field, the wave function becomes

$$|after\rangle = a(+1)|+1\rangle \psi_{+1}(x) + a(0)|0\rangle \psi_0(x) + a(-1)|-1\rangle \psi_{-1}(x) \qquad \text{(A2-2)}$$

where the subscripts indicate the three different spatial parts of the wave function take three different paths.

We now put a detector on each of the paths, and suppose there is an observer looking at the outcome (*yes* or *no*) displayed on each dial. The full wave function, particle plus detectors plus observer, is then the sum of three parts;

$$\begin{aligned}
|\,after\rangle = \\
a(+1)|+1\rangle\psi_{+1}(x)|D+1:yes\rangle|D0:no\rangle|D-1:no\rangle|Obs:yes,no,no\rangle + \\
a(0)|0\rangle\psi_{0}(x)|D+1:no\rangle|D0:yes\rangle|D-1:no\rangle|Obs:no,yes,no\rangle + \\
a(-1)|-1\rangle\psi_{-1}(x)|D+1:no\rangle|D0:no\rangle|D-1:yes\rangle|Obs:no,no,yes\rangle
\end{aligned} \quad (A2\text{-}3)$$

where the *a*'s are the 'sizes' of each option. So we see that the final state contains three separate versions of reality. In particular, there are three versions of the observer, each perceiving a different result. One and only one of the versions will correspond to what we, as 'the' observer, perceive, but quantum mechanics does not say which one. (But it does say that if the experiment is repeated many times, the +1 result will be observed on a fraction $|a(+1)|^2$ of the runs, and similarly for 0 and -1.)

**Spin ½.** There is one more intriguing point about angular momentum in quantum mechanics. For the hydrogen atom, the mathematics predicts that the *total* angular momentum can be 0, 1, 2, 3, 4,… and so on, with, respectively, 1, 3, 5, 7… *components* when measured along some axis. That is, there are an odd number of components (different states, different number of trajectories after the magnetic field). But when Stern and Gerlach measured the angular momenta of silver atoms, they found only *two* components, corresponding to an angular momentum of ½ (with allowed values + ½ and – ½ )! This, in fact, was how spin ½ was discovered.

Interestingly, the abstract group representational theory (with 'abstract' here meaning not associated with a particular problem) allows the observed spin ½, even though it does not occur in the hydrogen atom case. This may have theoretical implications (section 5, the last subsection, and reference [11]). Odd half integer spin particles ( 1/2, 3/2,…), known collectively as fermions, behave quite differently in certain situations from integer spin particles (0,1,2,…), known collectively as bosons.

# A3. Different Versions Are in Separate, Non-Communicating Universes.

We will show that the different versions of reality in the wave function are in separate universes and that there can be no communication between them. This result is demonstrated using a particular example, but the same arguments would hold whenever any 'single-particle' wave function (except perhaps that for a photon) simultaneously takes at least two different paths.

Fire a spin ½ silver atom into a Stern-Gerlach device. There will be one 'trajectory' traced out by the +½ wave functions and another traced out by the –½ wave functions. Let $\Omega_+(t)$ be the time-dependent three dimensional volume where the silver atom wave function is non-zero on the +½ trajectory and $\Omega_-(t)$ the corresponding volume on the –½ trajectory. These regions are well-defined even after the two branches of the silver atom wave function hit the detectors, and they are non-overlapping after the wave function of the silver atom clears the magnet.

Now consider the Schrödinger equation for the full wave function (including both versions of reality) describing the silver atom and the detectors;

$$i\frac{\partial(a_+\Psi_+(x_s,\{x_d\})+a_-\Psi_-(x_s,\{x_d\}))}{\partial t} = H(a_+\Psi_+(x_s,\{x_d\})+a_-\Psi_-(x_s,\{x_d\}))$$

(A3-1)

where the $a$'s give the relative sizes of the two versions, $x_s$ is the coordinate of the silver 'atom' and $\{x_d\}$ represents the coordinates of all the 'atoms' in the detectors (plus the observer if one wishes). $\Psi_+$ denotes the branch of the wave function when $x_s$ is in region $\Omega_+$, and $\Psi_-$ the branch when $x_s$ is in region $\Omega_-$. Because the two regions do not overlap, we have (assuming no long-range interactions for the silver atom)

$$\Psi_+(x_s,\{x_d\}) = i\partial_t\Psi_+(x_s,\{x_d\}) = H\Psi_+(x_s,\{x_d\}) = 0 \text{ when } x_s \text{ is in region } \Omega_-$$

(A3-2)

$$\Psi_-(x_s,\{x_d\}) = i\partial_t\Psi_-(x_s,\{x_d\}) = H\Psi_-(x_s,\{x_d\}) = 0 \text{ when } x_s \text{ is in region } \Omega_+$$

(A3-3)

(Note: The $H$ in $H\Psi_+(x_s,\{x_d\})$ does not (substantially) change the location of the silver atom. Thus $H\Psi_+(x_s,\{x_d\})$ in Eq. (A3-2) is a function of $x_s$ which has the value 0 when $x_s$ is not in $\Omega_+$.)

Eqs. (A3-1), (A3-2) and (A3-3) imply the wave functions for the two branches obey their own separate equations of motion,

$$i\partial_t\Psi_+(x_s,\{x_d\}) = H\Psi_+(x_s,\{x_d\}) \text{ (relevant when } x_s \text{ is in region } \Omega_+)$$

(A3-4)

$$i\partial_t\Psi_-(x_s,\{x_d\}) = H\Psi_-(x_s,\{x_d\}) \text{ (relevant when } x_s \text{ is in region } \Omega_-)$$

(A3-5)

(and the Schrödinger equation is irrelevant (0=0) when $x_s$ is in neither region.) Thus, because they obey their own separate equations, and because $H$ is independent of the wave function in a linear theory, the evolution of $\Psi_+$ is entirely independent of what happens on the $\Psi_-$ branch (and vice versa). That is, the different versions act as if they were in different, non-communicating universes.

# A4. Classical Consistency of Observations in Quantum Mechanics.

We wish to show here that quantum mechanics agrees with our perception of a seemingly classical world. The main ideas are that only versions of the observer perceive and that the versions of reality, including the versions of the observers, are in isolated, non-communicating universes.

To illustrate the principles, we consider an experiment on an atomic system with $K$ states, $|k\rangle$. There is an apparatus, denoted by $|A\rangle$ and an observer denoted by $|O\rangle$. After the experiment is done, the state is

$$\sum_{k=1}^{K} a_k |k\rangle |A_k\rangle |O_k\rangle \tag{A4-1}$$

Before the experiment starts, we tell the observer to write down "1, classical" if they see just one version of reality, and "other" if they perceive anything else (such as a "double exposure"). We see from Eq. (A4-1) that *each version $|O_k\rangle$ of the observer perceives only the $k^{\text{th}}$ outcome*, so each version will write down "1, classical;"

$$\sum_{k=1}^{K} a_k |k\rangle |A_k\rangle |O_k(1, classical)\rangle . \tag{A4-2}$$

"Other" is never written. Even if we switch bases, so that each version of the observer is a linear combination of the $|O_k\rangle$, "other" will never be written because $|O_k\rangle$ *in the linear combinations* is always $|O_k(1, classical)\rangle$. And since "other" is never written, it must never be perceived. Thus

> **2-6.** Perception of more than one outcome of an experiment by an observer can never occur within quantum mechanics.

Next we consider principle **2-9**, on the agreement of observers. We ask two observers to write down their perceptions and then ask the second observer to look at the first observer's result and write "agree" or "disagree." The resulting state is

$$\sum_{k=1}^{K} a_k \mid k \rangle \mid A_k \rangle \mid O^{(1)}{}_k \rangle \mid O^{(2)}{}_k, agree \rangle \qquad \text{(A4-3)}$$

Thus "disagree" is never written, so

> **2-9.** Quantum mechanics implies that two observers can never disagree on what they perceive.

We now consider principle **2-7**. We start from state (A4-1) and have the observer look twice at the dial. Then we have state

$$\sum_{k=1}^{K} a_k \mid k \rangle \mid A_k \rangle \mid O(k,k) \rangle \qquad \text{(A4-4)}$$

Thus we see we never have $\mid O(k,j) \rangle$ so that an observer never has conflicting memories:

> **2-7.** If two observations of a single result are made successively, quantum mechanics implies the same consistency of results as one obtains in a classical universe.

The same argument would apply if one did multiple experiments; results are always consistent within a version, and versions of the observer cannot 'see' from one isolated universe to another. That is, because the interactions are local and separable, a cause in the past can never produce a classically inconsistent effect in the future *within a version*. So each version of the observer perceives a classical cause and effect universe.

Finally, we separately consider principle **2-8**, on the consistency of successive measurements. As an example, we might have a spin ½ Stern-Gerlach experiment that yields two different possible trajectories for the particles exiting the apparatus. We then put a second Stern-Gerlach experiment in each exit beam to re-measure the z-component of spin. We will find that we get consistent results, both experimentally *and quantum mechanically*.

To actually write out the states to show this is a notational nightmare, so we will talk it through. We suppose there is a device, $M$, that splits the beam of incoming particles into $N$ different trajectories, and on each trajectory, there is a readout, R, that says *no* (no detection) or *yes* (detection). Then on each of the $N$ possible trajectories, we put a similar device, $M$ plus a similar readout. We designate the state of the apparatus before the experiment by a 0 subscript. Then for any of the $N+1$ sets of apparatus, we have

$$\sum a_k |k\rangle M |R_0\rangle \rightarrow \sum a_k M |k\rangle |R_k\rangle \qquad (A4\text{-}5)$$

and hence for the special case $a_j = 1, a_k = 0, k \neq j$

$$|j\rangle M |R_0\rangle \rightarrow M |j\rangle |R_j\rangle \qquad (A4\text{-}6)$$

where there is no sum in Eq. (A4-6).

Now we do the compound experiment. After the first apparatus, there will be *N* versions of reality. In version *k*, there will be the particle state, $|k\rangle$, plus a readout that says *k yes*, all others *no*. Now when the particle state $|k\rangle$ hits the second apparatus, *because of Eq.* (A4-6), that apparatus will also register *k* yes, all others *no*. Thus the second measurement must agree with the first, even if there is no collapse or no actual particle in a certain state. And so we have

> **2-8.** If one measures the same property twice in a row, quantum mechanics implies one will get the same result.

In summary, the perceptions of 'the observer' in quantum mechanics will never disagree with what is expected classically. (So the only problem with quantum mechanics is that we don't know which version of reality will pop up.)

## A5. Problems with Probability in the Many-Worlds Interpretation.

The probability law is about the probability of perceiving a given version of reality. In section 4, we argued that there could be no probability of perception in QMA (no particles, no collapse, no sentient beings) because (1) only versions of the observer perceive and (2) each version of the observer perceives its associated version of reality with 100% probability. Thus there is nothing probabilistic in QMA, which implies there can be no probability law.

In this appendix, because this result is important, we will present the argument against probability in QMA using two somewhat different lines of reasoning. In the first, we argue that one cannot properly *state* the probability law in QMA, so it certainly cannot hold there. And in the second, we show that the probability law implies that a version must be *singled out* for perception. As in the text, we argue from the state

$$|\Psi\rangle = \sum_{i=1}^{n} a(i) |i\rangle |\det(i)\rangle |obs(i)\rangle \qquad (A5\text{-}1)$$

so there is an equally valid version of the observer perceiving each outcome, with no version is singled out as *the* observer.

**The impossibility of stating the probability law in QMA.** It would seem to be straightforward to state the probability law. One possibility is:

**P1**. If many runs of the experiment are made, the probability that the observer will perceive outcome $i$ is $|a(i)|^2$.

This statement is certainly correct experimentally, but it is not acceptable within the framework of QMA. "…the observer will perceive" implies perception by a *unique* version of the observer—but there is, in QMA, no unique, singled-out version. Instead there are $n$ equally valid versions, each perceiving a different result. So **P1** will not do.

A second potential statement is:

**P2**. The probability of perceiving outcome $i$ is $|a(i)|^2$.

But this dodges the issue of what it is that perceives, and that is not acceptable in this context.

A third possibility, which acknowledges that it is the versions which perceive, is:

**P3**. On a given run, the probability of my perceptions corresponding to those of version $|obs(i)\rangle$ is $|a(i)|^2$.

But because there is a version of *me* associated with *every* outcome, "*my* perceptions" are not uniquely defined. When every outcome is perceived by a valid version of me, there can be no probability associated with *my* perceptions. Thus **P3** will also not do in QMA (although it is quite acceptable 'in reality').

We might try:

**P4**. The probability of $|obs(i)\rangle$ perceiving outcome $i$ on a given run is $|a(i)|^2$.

But this won't do because $|obs(i)\rangle$ perceives outcome $i$ on *every* run.

Finally, we might try:

**P5**. The probability of outcome $i$ occurring is $|a(i)|^2$.

But that is not acceptable either, because all outcomes *occur* on every run so the only probability relevant to QMA is probability of *perception*.

The point is this: The probability law is about what is perceived, and the only entities that perceive in QMA are the $n$ versions of the observer. So

*The probability law in QMA must be written solely in terms of the perceptions of the versions of the observer,*

with no reference to "*my*" perceptions or the perceptions of "*the*" observer (because there is no "*me*" or singular, unique observer different from the versions). But that seems patently impossible because every version of the observer perceives its respective outcome on every run with 100% certainty; there is nothing probabilistic about the perceptions of the versions.

To restate the argument: When only the versions perceive, when every version is of equal perceptual status so that no version is singled out as 'me,' and when each version perceives its respective outcome on every run—that is, when there is no probabilistic or coefficient-dependent aspect to the *n* perception processes—there is no way within QMA to obtain a coefficient-dependent probability of perception.

We note that the statements **P1** and **P3**, acceptable in reality (but not in QMA), are most easily understood if one version of the observer is somehow singled out (in a valid interpretation) as being *the* observer.

**Adding the probability law as an assumption.** Suppose we try an interpretation, QMB, which consists of QMA plus the *assumption* that the probability law holds. This doesn't work either because, under the restrictions of QMA (no particles, no collapse, no sentient beings), as we have observed, one cannot *state* the probability law. Thus the probability law cannot simply be added on because, when only the versions perceive, the law cannot be properly stated.

Or, we could say that it's permissible to add the assumption, but this new QMB interpretation comes with baggage; it must contain some mechanism for singling out one version as the one corresponding to "*me*." If one version is not singled out, the probability law cannot hold because there is no probability of perception of the versions in QMA. (Also, see below, on the necessity for singling out.)

**Other approaches to probability in the many-worlds interpretation.** There have been a number of attempts [36-41] to introduce probability into QMA through a 'subjective' process: After the experiment is done and the result recorded by the versions of the apparatus, but before the observer perceives the recorded results, each version of the observer is uncertain about what he will perceive. This *uncertainty* is said to lead to a *coefficient-dependent probability* for *expectations* of perceptions. But no logical explanation has been given—from within the confines of the QMA scheme rather than from the perspective of 'actuality'—for why uncertainty implies a *coefficient-dependent* probability. In fact, one can give an analogy which shows that the size of the coefficients is irrelevant in this situation. Suppose *n* observers are led, blindfolded, into *n* different-sized rooms (analogous to the *n* different branches with different norms). If one of the observers is asked the probability of being in room 3, surely the size of the rooms has nothing to do with the answer.

In addition, the subjective method seems irrelevant to the problem of stating the probability law in QMA because the experimentally observed law must be stated in terms that pertain to *actual perceptions* rather than the *expectations* of the perceptions by the versions.

There have been attempts to *derive* the probability law [38, 42-44] from within QMA but none of them has stated the law in an acceptable form. Unless one can suitably state the probability law, which I claim is impossible, the conclusion is that the explicit or implicit assumptions made in these attempted derivations—for example, that probability is a property of the wave functions—carry one beyond the confines of the un-amended QMA.

**The necessity for singling out.** By looking at a particular case, we can see the problem with probability in QMA from a different perspective. Suppose we consider a two-state system,

$$\psi = a_1 |1\rangle + a_2 |2\rangle, \tag{A5-2}$$

with $|a_1|^2 = .9999, |a_2|^2 = .0001$, and suppose we do ten runs, with the observer perceiving the results of every run. There will be $2^{10} = 1,024$ possible outcomes and 1,024 versions of the observer, each equally valid. But we know experientially that my perceptions will (almost always) correspond to only one of them, the one with all ten outcomes 1. That is, one version from among all the 1,024 versions is (almost always) *singled-out* as the one corresponding to my experiential perceptions.

But in the QMA interpretation, every one of the 1024 versions of the observer perceives its respective outcome on every run of 10. So why am "I" never the version of the observer that perceives, say, five outcomes 2? There is no reason *within QMA*, where all versions of me are of equal perceptual status (and there is no unique, 'external' me separate from the versions). That is, our awareness of only the ten outcomes 1 state is not consistent with the bare QMA.

More generally, if *I* consistently perceive one version more than another, then there must be a unique version singled out as *me* on each run. (One could counter that all versions are equivalent, but the bias comes in the assignment of "*my*" perceptions to one version or the other. But in this case, one has still assumed a singular, unique "*me*," different from the versions.)

The seemingly inescapable conclusion is that there must be a coefficient-dependent mechanism (collapse? hidden variables? a "Mind?"), outside QMA, which singles out one version of the observer as the one corresponding to "my" perceptions.

# A6. All Versions of the Observer Are Equally Aware.

When we perceive the results of an experiment with many potential outcomes, we are consciously aware (however you wish to define that) of one and only one outcome. So we might expect that one and only one version of the observer is aware in quantum mechanics. But we will show here that that is not the situation in the QMA (no particles, no collapse, no sentient beings) interpretive scheme; all versions are *equally aware*.

From the point of view of neuroscience or philosophy, what is meant by awareness or consciousness is not so easy to pin down. But we can define it in a general way that is suitable for our purposes here. The wave functions are all that exist in QMA, and so awareness can only correspond to some property—synchronous oscillations of many regions of the brain, for example—of the wave function of the observer's brain or brain-body.

To see what happens to the observer's awareness when a measurement results in several simultaneously existing versions of the observer, we consider an experiment on an atomic system with $K$ states, $|k\rangle$. There is an apparatus, denoted by $|A\rangle$ and an observer denoted by $|O\rangle$. The initial state, before the measurement, with an aware observer, is

$$|O_0(\text{aware})\rangle |A_0\rangle \sum_{k=1}^{K} (a_k | k\rangle) \qquad (A6\text{-}1)$$

We do the measurement in two steps. First we let the apparatus measure and record the results but we cover the readout of the apparatus with an X, so the state is

$$\sum_{k=1}^{K} (|O_k(\text{aware of X})\rangle |A_k\rangle |k\rangle a_k). \qquad (A6\text{-}2)$$

where the subscript on the observer state indicates that that version is in universe $k$. Now we have 'the observer' look at the reading on the apparatus. We suppose that some versions of the observer are aware and others are not. Then the final state is

$$\sum_{k,\,aware} (|O_k(\text{aware of k})\rangle |A_k\rangle |k\rangle a_k) +$$

$$\sum_{k,\,not\,aware} (|O_k(\text{not aware})\rangle |A_k\rangle |k\rangle a_k) \qquad (A6\text{-}3)$$

But the $K$ states are in separate, non-communicating universes (see appendix A3), and there can therefore be *no coordination* between the evolutions of the versions of the observer from the aware state of (A6-2) to either an aware or a not-aware state in (A6-3). The evolution of each version to an aware or not-aware state is *individually, separately* determined by 'chance,' perhaps depending on the way the photons hit the eye in the different versions. This means that, if the not-aware sum is non-zero—if an aware state of the observer is sometimes carried by chance to an unaware state—then by chance (because there is no coordination between versions) there would sometimes be a situation in which *no* version of the observer was aware at the end.

But from our experience of awareness, that is not an allowable outcome; we are always aware of a result. So the only way the time evolution can be made consistent with the at-least-one-aware requirement is to suppose the observer state $|O_k(\text{aware of X})\rangle$ evolves to an aware version, $|O_k(\text{aware of k})\rangle$, *on every branch*. *All versions are equally aware* in QMA, just as aware as the original version of the observer! Thus if "*I*" perceive one result, there are $n-1$ other, equally aware "*I*"s perceiving the other results.

Comment 1: Instead of 'awareness,' we could use any property that might be used to single out just one version and the argument still goes through. In particular, even though we are *experientially* aware of just one outcome, this result implies that no one version can be singled out as *me* in QMA; there are *K* equally valid versions of me.

Comment 2: There is one way the above argument could fail and that is if the awareness-distinguishing feature is carried by the apparatus or the atomic system. Then the distinguishing feature could be transferred to the observer wave function in the transition from (A6-2) to (A6-3). But a singling-out of this sort amounts to a hidden variable theory (one version 'physically' distinguished from the others), and that is forbidden by hypothesis here.

# A7. Small Parts of the Wave Function Carry the Full Energy, Momentum and so on.

One of the primary reasons Einstein proposed his particulate photon model of light (in 1905, before quantum mechanics) was that a classical light wave could not transfer a sufficient amount of energy to electrons to knock them out of a metal. The reasoning was that a small portion of the (classical) light wave carried a correspondingly small part of the energy, and since an electron was small, it could not quickly accumulate enough energy from the light wave. But with a proper understanding of the properties of the wave function, we can see that Einstein's reasoning is not applicable.

The basic idea is that when a wave function splits into several parts, each part carries the *full* mass, energy, momentum, spin, and charge. This holds because the operators associated with these quantities are linear and *local*. We will illustrate for energy and momentum, but the argument would be similar for the other quantities.

We shoot a spin ½ particle into a Stern-Gerlach apparatus. Before the particle reaches the magnet, the wave function is

$$\psi(x,t) = \int d^3k \, f_{k_0}(k) e^{ik \cdot x - i\omega(k)t}, \quad \int d^3k \, |f_{k_0}(k)|^2 = 1 \qquad \text{(A7-1)}$$

where $|f_{k_0}(k)|$ is sharply peaked about $k_0$. Then to a good approximation,

$$\begin{aligned} i\hbar \partial_t \psi &= \hbar \omega(k_0) \psi = E\psi \\ -i\hbar \partial_x \psi &= \hbar k_0 \psi = p\psi \end{aligned} \qquad \text{(A7-2)}$$

so that the initial wave function is essentially an eigenfunction of energy and momentum.

Now the wave function goes through the magnet and splits into two parts,

$$\begin{aligned} \psi(x,t) &= a_{+1/2} \psi_{+1/2}(x,t) + a_{-1/2} \psi_{-1/2}(x,t), \quad |a_{+1/2}|^2 + |a_{-1/2}|^2 = 1 \\ \psi_{+1/2}(x,t) &= \int d^3k \, f_{k'_0}^{+1/2}(k) e^{ik \cdot x - w(k)t}, \quad \int d^3k |f_{k'_0}^{+1/2}(k)|^2 = 1 \\ \psi_{-1/2}(x,t) &= \int d^3k \, f_{k''_0}^{-1/2}(k) e^{ik \cdot x - w(k)t}, \quad \int d^3k |f_{k''_0}^{-1/2}(k)|^2 = 1 \end{aligned} \qquad \text{(A7-3)}$$

where the primed and double-primed $k$'s indicate where the corresponding $f$'s have a sharp peak, with $k'_0 \cong k''_0 \cong k_0$. But we see that

$$\begin{aligned} i\hbar \partial_t a_{+1/2} \psi_{+1/2} &= \hbar \omega(k'_0) a_{+1/2} \psi_{+1/2} = E a_{+1/2} \psi_{+1/2} \\ -i\hbar \partial_x a_{+1/2} \psi_{+1/2} &= \hbar (k'_0) a_{+1/2} \psi_{+1/2} = p a_{+1/2} \psi_{+1/2} \\ i\hbar \partial_t a_{-1/2} \psi_{-1/2} &= \hbar \omega(k''_0) a_{-1/2} \psi_{-1/2} = E a_{-1/2} \psi_{-1/2} \\ -i\hbar \partial_x a_{-1/2} \psi_{-1/2} &= \hbar (k''_0) a_{-1/2} \psi_{-1/2} = p a_{-1/2} \psi_{-1/2} \end{aligned} \qquad \text{(A7-4)}$$

So each of the two separate parts has (nearly) the same energy and momentum as the original wave. The energy and momentum did *not* split up according to the values of $|a_{+1/2}|^2$ and $|a_{-1/2}|^2$. This means the small part of the light wave function that hits one electron in the photoelectric (or Compton) effect carries the full energy and momentum (and spin) of the wave function.

**Conservation Laws.** One might imagine that if each small part carries the full energy and momentum, we have given too much energy and momentum to the full wave function. But energy and momentum (and charge, etc.) do not add across sums; they only add across *products* of wave functions (because the corresponding operators are essentially first-order derivatives).

**Classical Electromagnetic Waves.** Why does a classical electromagnetic wave behave differently in this respect from the wave function? Because it is made up of many *localized* light-like wave functions, and each small portion of the classical wave contains only the energy and momentum of those photon-like wave functions localized in that region of space.

# A8. Bell's Theorem and Non-Locality.

In 1964, Bell [12] proved that if all the information regarding a particle-like state was contained locally, say within a centimeter of the localized wave function, then there would be a conflict with quantum mechanics. In particular, he showed that if two particles, electrons or photons, were initially in a spin 0 state which split into two single-particle states that moved away from each other, there would be correlations between the two measured spin states that would be different from what quantum mechanics predicted. This experiment has been done by Aspect [13] and others, and it was found that the quantum mechanical laws were obeyed. Thus we have a proof that there can be no *local* hidden variable theories.

Our view of these results is: In interpreting the results of this experiment, is often assumed there are (localized) particles. Under that assumption, one needs instantaneous long-range signaling between the particles to account for the correlations in the Aspect experiment. (And this is often stretched to imply that everything is interconnected.) But if there are no particles (section 7), there is no need for signaling; it is just quantum mechanics as usual, with its inherent non-locality.

**Non-locality via the wave function.** With all due respect to Bell—he started a major line of inquiry in quantum mechanics—it seems to me that one could reasonably expect hidden variable theories to be non-local, in contrast to Bell's assumption. First, one knows that the wave functions of quantum mechanics contain non-local information; the Aspect experiment and theory tell us that. Second, we see from the Bohm hidden variable model (page 22) that the hidden variables can depend locally on the wave function (velocities are derivatives of the local phase angle of the wave function). So one could imagine a hidden variable theory in which there is indirect non-locality: Each of the hidden variable sets for the two particle-like states takes its information *locally* from the *wave function*. But because the wave function contains non-local information, the conditions for Bell's theorem no longer apply. Thus there are reasonable objections to Bell's locality assumption for hidden variable theories.

**Correlations vs. 'influence.'** On the surface, it looks like the measurement on one 'particle' influences the state of the distant 'particle.' But if there are no particles, and if there is no collapse, then detection of the state of one particle-like wave function does not in any way affect or influence the quantum state of the second, distant particle-like wave function. All we can say is there is a correlation between the measurements on the two distant particle-like wave functions.

## A9. The Wheeler Delayed-Choice Experiment.

A Mach-Zehnder interferometer is set up so that a light wave is divided by a half-silvered mirror and travels on two different paths. After the photon-like wave function has been divided but before it is detected, the detector at the end of the interferometer is rapidly adjusted to one of two possible settings. If particulate photons exist, then the measured results from the two different settings depend on which path the conjectured particulate photon took. Under this assumption, the experimental results show that the conjectured photon made a choice of path *after* it had passed the point where the paths divide—a violation of our intuitive understanding of causality.

If, however, one assumes only the wave function exists (no particulate photons), then quantum mechanics, as it is, gives the correct answer, with no after-the-fact choice involved. That is, the assumption of the existence of particles gets us in causality trouble, but no-particle quantum mechanics does not. So I would take the results of the Wheeler delayed-choice experiment [14,15] as more evidence (in addition to that of sections 5, 6, and 7) for no particles rather than as causality operating backwards in time (from the later time when the detector setting was changed, with the photon was halfway to the second mirror, to the earlier time when the photon had not yet passed the first mirror).

## A10. The Quantum Eraser Experiment.

The aim of this set of experiments [16,17] is similar to that of the Wheeler delayed-choice experiment in that *if one assumes* the existence of particles, then one gets into causality—and locality—troubles. Before explaining the experiments, we should say that no-particle quantum mechanics gives results in agreement with the experiments. So our view is that one is using an unsupported idea—particles—to derive seemingly surprising results.

**Experiment 1.** The experiment uses two entangled photons, originally in a spin 0 state, that fly apart from each other. If one has polarization in the *x*-direction, then the other has polarization in the *y*-direction, no matter what orientation is chosen for the *x* and *y* axes. In this first experiment, photon 1 goes through a double slit and is then detected on a screen while photon 2 is simply detected. The results are not surprising; when the experiment is run many times, one gets the usual **double-slit** interference pattern on the photon 1 screen.

**Experiment 2.** The light from the two slits in photon 1's path are put through special crystals that tinker with the polarization (see appendix A1) so that the polarization of the light that goes through one slit is completely different from the polarization of the light that goes through the other slit. In that case, the light from the two different slits cannot interfere, so one gets a **single-slit** interference pattern on the screen.

**Experiment 3.** The light from *photon 2* is now *polarized* and detected. The polarization of the second photon has the effect that the crystals in front of the two slits no longer give different polarizations to the light from the two different slits. Thus the two beams interfere and one again gets a **double-slit** interference pattern spread out over the screen.

From one point of view, this seems remarkable. The experimental arrangement for photon 1 was not changed from experiment 2 and yet the results for photon 1 are different from those of experiment 2. This would *appear* to be saying that if there are particles which are in a definite state—position and polarization—at each instant, then the polarization state of particle 2 affects the trajectory (from slits to screen) of the distant particle 1.

**Experiment 4.** Same experimental setup as in experiment 3 except that the polarizer and the detector for photon 2 are moved far away, so photon 1 is detected before photon 2. One still gets a **double-slit** interference pattern. (But if photon 2 were not detected, or if it were detected as in experiment 2, one would get a single-slit pattern.)

Again from a certain point of view, this is saying that the state of particle 2, determined only after particle 1 is detected, *retroactively* affects the trajectory of distant particle 1. That is, causality appears to work backwards in time.

**Explanation.** In all the experiments, there is a coincidence circuit that accepts only those results where both detector 1 and detector 2 detect (within a certain time frame). So in experiments 3 and 4, one is measuring the pattern made by many 1 photons *given* that photon 2 has a certain polarization. If the polarizer for photon 2 were rotated 90 degrees, one would still get a double-slit interference pattern, but if the two double-slit interference patterns for 1 ( from 0 degrees and 90 degrees) were superimposed, they would just give the single-slit pattern of experiment 2! Thus the polarizer in path 2, plus the coincidence counter, effectively *selects* for a certain set of 1 photons and these give the double-slit pattern even though the set of *all* 1 photons gives the single-slit pattern.

If one postulates particles, and if one requires that each particle be in a definite state at each instant, then experiment 3 seems to require action-at-a-distance between the two particles. And experiment 4 seems to require *retroactive* action-at-a-distance. But if one postulates no-particle quantum mechanics, the experiments simply verify the *correlations predicted by quantum mechanics* between the two entangled photon-like wave functions.

This is a summary of the write-up at .grad.physics.sunysb.edu/~amarch which is best reached by Googling "quantum eraser."

# A11. Problem with the Kochen-Specker Proof of No Hidden Variables.

It has been conjectured that even though one cannot show there are particles, there may still be 'hidden' (not detectable by experiment) variables that determine the unique outcome of an experiment, or at least the outcome we perceive. At a given time, there is a unique, single-valued set of these hidden variables (as opposed to the many versions of reality of the wave function), so they are objective.

**The Kochen-Specker reasoning.** Kochen and Specker [22], and later Conway and Kochen [23], used an ingenious argument in an attempt to show that it is not possible to have an underlying hidden variable theory. The primary characteristic of a hidden variable theory is that,

**A11-A.** Once the hidden variables (whatever they are) are set, the outcome of every possible experiment is determined.

Kochen and Specker then showed that for certain sets of experiments, this led to a contradiction. That is, they claim it is logically impossible to satisfy principle **A11-A** in every situation! If this is correct and free of hidden assumptions, then it would constitute a proof that there could be no hidden variable theory.

To explain their reasoning in more detail, we will review the Conway-Kochen paper arXiv 0604097, *The Free Will Theorem* [23]. The system investigated has spin 1. In such a system, the three components of spin squared, $\sigma_x^2, \sigma_y^2, \sigma_z^2$, commute, so they can be simultaneously measured. The possible outcomes are 0 and 1, and it can be shown that only one of the three can give 0, so that the other two must yield 1. It is assumed that the hidden variables $\lambda$ are set, so that the results of a set of 3 measurements along *any* set of axes are determined. They then find a series of sets of 3 measurements such that one eventually ends up with two measurements along orthogonal axes both giving 0, which is a mathematical impossibility.

On the other hand, we have the Bohm model, which is a valid hidden variable theory within its limitations. That is, it is constructed in such a way that, for any experiment, the $|a(i)|^2$ probability law is always satisfied. We can construct a spin 1 system in the Bohm model by having two spin 0 particles in a bound state. And there is no overt requirement in the K-S-C method, as far as I know, that disqualifies the Bohm model. (The Bohm model is not relativistic, but relativity does not enter the K-S-C impossibility argument.) From this we are led to conjecture that there must be some hidden assumption in the K-S-C reasoning which makes it invalid.

The most likely problem is that one can show that **A11-A** is wrong. To see why, consider the bound-state spin 1 system in the Bohm model. In the model, one derives trajectories for each of the particles from the Schrödinger equation, and it is imagined that particles are put on one of those two-particle trajectories. We will call the hidden variables, the positions and velocities of the two particles, specified at time 0, $\lambda_0$. Then the two particles follow very complicated trajectories. Suppose at some point in time, we measure the z spin component squared. Because of the complicated trajectories, fixing $\lambda_0$ does not fix the measured value; that value also depends on *when* (using time-of-flight after fixing $\lambda_0$) the measurement is made. And the oscillation in time between different outcomes is extremely rapid. Thus principle **A11-A** should read

> **A11-B.** Once the hidden variables *and the detector geometry* are set, the outcome of every possible experiment is determined.

But because K-S-C have not specified the experimental setup for measuring the components of spin, they have not taken the detector geometry into account in their argument. Because the geometry changes as one measures along different sets of axes, it is quite possible that their assumptions about how measurements carry over from one set of 3 measurements to another is incorrect.

So before one can accept the K-S-C "no-go" theorem, it needs to be shown that one can ignore the role of the detector geometry. Further, one needs to know how to deal with the apparent counter-example, the Bohm model, to their no-go theorem.

# A12. Mathematical Collapse Implies Non-Linearity.

The primary characteristic of quantum mechanics is that it is linear. Therefore it is logical to ask whether mathematical collapse can fit within a linear theory in which the probability law holds. We find that it cannot; the probability law requires a mathematical collapse to be non-linear.

**Non-unitary time translation.** If the Hamiltonian *H* is Hermitian (so time translation is unitary), as it is in conventional theory, then the norms of the different branches of the wave function stay the same forever. So there can be no collapse if one sticks to a linear, *unitary* theory. But what would happen if we kept the Hamiltonian linear but allowed it to be non-Hermitian so that time translation is no longer unitary? Suppose in particular that some part of the Hamiltonian can change the norm of the wave function, with the change depending on the random variables *w*, where the choice of *w*'s at each stage in the time evolution is independent of the coefficients (linearity). Then we have for the general linear case, with *O* being the time translation operator,

$$\Psi(t,w) = O(t,w)\sum_{k=1}^{K} a_k \Psi_k(t=0)$$
$$= \sum_{k=1}^{K} a_k O(t,w) \Psi_k(t=0)$$
$$= \sum_{k=1}^{K} a_k \beta_k(t,w) \Psi_k(t=0) \qquad \text{(A12-1)}$$
$$= \sum_{k=1}^{K} \alpha_k(t,w) \Psi_k(t=0)$$

where the conjectured Hamiltonian is presumed to have no effect on the 'shape' of the wave function (and we have ignored the non-collapse part of the Hamiltonian). In this case, it might happen that after some time has elapsed, for some value of *i*,

$$\frac{\beta_i^2(t,w)}{\sum \beta_k^2(t,w)} \to 1, \ t \to \infty \qquad \text{(A12-2)}$$

so we end up with a system which has collapsed to state *i*. Thus linearity does not rule out collapse.

**The $|a(i)|^2$ probability law requires collapse to be nonlinear.** But now suppose the collapse follows the $|a_i|^2$ probability law. Then for large *t* and given *i*, Eq. (A12-2) will hold (depending on the sequence of *w*'s chosen) a fraction $|a_i|^2$ of the time and the same equation with 1 replaced by 0 will hold the rest of the time. Thus to satisfy the $|a_i|^2$ probability law, the average (averaged over the choice of *w*'s) of the ratio must obey

$$\langle \frac{\beta_i^2(t,w)}{\sum_{k=1}^{K} \beta_k^2(t,w)} \rangle = \int dw P(w) \frac{\beta_i^2(t,w)}{\sum_{k=1}^{K} \beta_k^2(t,w)} = |a_i|^2, \ t \to \infty \qquad \text{(A12-3)}$$

where *P* is the probability of a certain set of *w*'s, and the integral over all possible *w*'s is suitably defined. If *P(w)* is independent of the coefficients, this cannot hold because the left hand side has no dependence on the $a_i$. Thus to satisfy Eq. (A12-3), $P(w)$ must depend on the coefficients.

Suppose now we think of numerically integrating the equation of motion, including the $w$-dependent random part, step by small step. At each step in the integration, the choice of $w$'s dictated by $P(w) = P(w; \alpha_1, ..., \alpha_K)$ depends on the coefficients. And then the dependence of the Hamiltonian on the $w$'s gives an indirect dependence of the Hamiltonian on the coefficients, so that different initial $a_i$'s will produce different $w$'s. But then

$$\Psi(t, w) = O(t, w) \sum_{k=1}^{K} a_k \Psi_k(t=0) \neq$$
$$\sum_{k=1}^{K} a_k O(t, w) \Psi_k(t=0) \qquad (A12-4)$$

and so the equations of motion are non-linear, in contrast to current quantum mechanics.

## A13. Collapse by Conscious Perception.

To illustrate the difficulties facing the proposal of collapse by conscious perception, we will look at a Stern-Gerlach experiment on a spin ½ silver atom. There is a detector on the trajectory of the – ½ branch but none on the trajectory of the + ½ branch, with the detector reading "yes" for detection and "no" for no detection. Suppose the observer is initially gazing off into space and then, a couple of seconds after the silver atom passes the detector, she focuses her attention on the detector.

Presumably, for an instant, there are two versions of the observer, perceiving *both* the yes and the no branches. In this instant, there must be something—we will call it a "Mind"—outside quantum mechanics, which realizes 'the observer' is perceiving a multi-version reality. That is, the "Mind" must somehow become aware that there are two separate versions of the observer's brain wave function. And then, in the next instant, the "Mind" collapses the wave function to just one option.

There are three problems with this approach. They do not rule it out, but they still need to be addressed. The first is that conscious perception—focusing the eyes on the dial in this case—is a complicated neural process. So it is not clear at what point in that process the "Mind," perceiving the state of the brain, knows there are two versions.

For the second problem, suppose the observer focuses on the dial the whole time, from before the atom is shot through the apparatus to after it passes the detector. And suppose the final result is that the detector is perceived as continuing to read no, so there is, in the conventional sense, no change in the dial reading from beginning to end. In that case, the observer has not *consciously* (in the conventional sense) perceived anything that could have cued the collapse. This implies the definition of "conscious perception" must be amended from its conventional meaning to accommodate collapse by "conscious perception." It also implies, as indicated above, that the Mind, associated with the individual brain, is perceiving more than one version.

To illustrate the third problem, we again use a Stern-Gelach experiment, with detectors on both branches. The state is

$$a(+1/2)\,|+1/2\rangle\,|\det(+1/2),yes\rangle\,|\det(-1/2),no\rangle\,|\,obs\,brain\ registers\ yes,\ no\rangle +$$
$$a(-1/2)\,|-1/2\rangle\,|\det(+1/2),no\rangle\,|\det(-1/2),yes\rangle\,|\,obs\,brain\ registers\ no,\ yes\rangle$$

$$(A13\text{-}1)$$

Now each version of the observer's brain is independent of the coefficients; for example, the neural firing pattern for each version does not depend on $a(+1/2)$ or $a(-1/2)$. However, the "Mind" must collapse the wave function in accord with the $|a(i)|^2$ probability law, so the "Mind" must have access to the values of those coefficients. But it is not possible to get those values strictly from the wave function of the observer's brain. So some process must be proposed for how the "Mind" 'senses' the value of the coefficients. This observation virtually guarantees that the collapsing "Mind" must perceive more than just the wave function of the individual brain. To obtain information on the coefficients, it must apparently perceive the whole wave function.

# A14. Spread of the Wave Packet. Application to Brain Processes.

The question is whether quantum mechanics is relevant in the brain. The primary dynamics of the brain is that there is an electrochemical pulse which travels from one end of a neuron to another. The propagation of this pulse is classical; quantum mechanics is irrelevant here (as far as we currently know). But the *initiation* of pulses makes use of the synapses, the small separations between adjacent neurons, and these synapses are small enough—less than a thousand atoms or 100 nm across—that quantum processes are potentially relevant. In fact, using an argument due to Stapp [4], we can show that quantum considerations are almost certainly relevant for neural processes.

**Synapses and calcium ions.** The synapses determine whether or not a pulse in one neuron triggers a pulse in the next neuron, and so they control the flow of thoughts. The passing-on or not-passing-on is in turn partly controlled by calcium ions. Classically and simplistically speaking, if a calcium ion is near that edge of a synapse which is next to the synapse of another neuron, •) (, the pulse will be passed on. That is, the positions of the calcium ions, in effect, control the flow of thoughts.

Now the wave function of a calcium ion (or any particle) doesn't just sit, unmoving, in space; it spreads out. So one way that quantum mechanics can be relevant to the brain is if a single calcium ion spreads out over the whole synapse in a reasonable time. If it does, within a time appropriate for neural processes, then the *quantum* state of the synapse will be a linear combination of [passing on] and [not passing on] the electrochemical pulse. In that case, quantum mechanical considerations are indeed relevant to the functioning of the brain.

**The spread of calcium ions.** So the question becomes: Do calcium ions spread out over the synapse in the time, approximately 1 millisecond, relevant to neural processes? To see, we use the uncertainty principle

$$\Delta x \Delta p \approx \hbar \tag{A14-1}$$

Suppose we have a calcium atom initially centered at the origin and use

$$\Delta x = x, \Delta p = mv = mdx/dt \tag{A14-2}$$

From these two equations, we get

$$xm\frac{dx}{dt} = \frac{m}{2}\frac{d(x^2)}{dt} = \hbar \Rightarrow x^2 = x_0^2 + 2t\frac{\hbar}{m} \tag{A14-3}$$

where $x_0$ is the initial spread in the wave function.

A reasonable initial spread for the wave packet is 10 nm, and a conservative representative time for brain processes is a tenth of a millisecond. We also have $\hbar \approx 10^{-34}$ and $m_{Ca} = 66x10^{-27}$ in mks units. If we put these numbers in, we get $x_{spread} \approx 500$ nm, which is more than adequate for the calcium ion to spread around the synapse.

The conclusion is: Because the wave function for a single calcium ion quickly spreads around the synapse, the quantum state of the synapse is a linear combination of passing on and not passing on the pulse. Thus quantum mechanical considerations certainly cannot be ruled out for brain processes.

One further point. There is a book [45] by the Nobel laureate brain researcher John Eccles called "*How the SELF Controls Its BRAIN.*" Because of the spread of the calcium wave packets, the wave function of the brain is a linear combination of trillions of different possible thoughts and signals for bodily actions. The Mind, if one subscribes to that interpretation, can 'control' the thoughts and actions by choosing one of those neural states to concentrate on. But the organizational problem, how to pick out just the right combination of firing neurons to cause a particular thought or bodily movement, is formidable. It must presumably be solved by some combination of the structure of the physical brain along with a set of organizational *templates* that are indigenous to the 'non-physical' Mind.

# A15. Comments on Probability in the Mind-MIND Interpretation.

**Non-Standard Probability.** How could the perceived results of many runs of the Stern-Gerlach experiment always be near $m/N=|a_1|^2$ if just the end result is perceived, but not near that value if every intermediate result is perceived? It does *not* arise because the non-physical Mind is influencing the outcome of each event (there is, in fact, no 'outcome of each event' because all the versions of reality continue forever). To see what is happening, suppose the 'probability law' for the Mind perceiving event $i$ is $s(x) = x + .1(x)(1-x)(.5-x)$ where $x$ is the amplitude squared. Then if each individual result is observed, the value for $m/N$ will be near $s(x)$, $x = |a_1|^2$, in disagreement with the $|a_i|^2$ probability law. But if only the end result is observed, the value for $m/N$ will be the one that maximizes $s(x)$ where $x$ is the appropriate amplitude squared,

$$x = \frac{N!}{m!(N-m)!}|a_1|^2|a_2|^2 \tag{A15-1}$$

From Stirling's approximation for the factorial and the chain rule for derivatives, we find this gives a maximum at $m/N = |a_1|^2$, in agreement with the $|a_i|^2$ probability law.

**The Rutherford Experiment.** One might suspect that experiments in which every outcome is observed have already been done. In particular, suppose we consider the original (1911) Rutherford scattering experiment, where a flash of light was observed (by a graduate student using a microscope) every time a decay particle hit the zinc sulfide detector. This experiment won't do as a test of the probability law, however, because *not every outcome* is observed. Only that very small fraction of events where the decay particle hits the small detector is observed.

And one can show in that case that the $|a_i|^2$ law will indeed hold (except perhaps for an overall normalization factor) no matter what the 'probability law' is for the perception of individual events.

To sketch the argument, suppose we do a scattering experiment in which there are R scattering events per second, but we observe only those events which register in a very small solid angle. Then one can prove that the scattering amplitudes squared are, to a good approximation, functions of the product $Rt|a_i|^2$ rather than being functions of $|a_i|^2$ and $Rt$ separately, where $t$ is the time the experiment has been run, and $a_i$ is the amplitude for scattering into solid angle $i$. This implies that, *independent of the functional dependence of probability on $|a_i|^2$*, the average time for observing a single event $i$ is proportional to $1/(R|a_i|^2)$, which is exactly the result one obtains if the probability law is $|a_i|^2$. Thus the individually observed scattering events in the original Rutherford experiment give no information on the specific probability law.